\documentclass[12pt]{article}
\usepackage{epsf}
\begin{document}
\newcommand {\sheptitle}
{Normal versus inverted hierarchical models within $\mu$-$\tau$ 
  symmetry}
\newcommand {\shepauthor}
{N.Nimai Singh$^{\dag,*}$\footnote{Regular Associate of ICTP.
\\ {\it {E-mail address}}: nimai03@yahoo.com}, H. Zeen Devi$^{\dag}$ and Mahadev Patgiri$^{\ddag}$}
\newcommand{\shepaddress}
{$^{\dag}$ Department of Physics, Gauhati University, Guwahati-781014, India\\
$^{\ddag}$ Department of Physics, Cotton College, Guwahati-781001,
India\\
$^*$ The Abdus Salam International Centre for Theoretical Physics,
Strada Costiera 11, 31014 Trieste, Italy. }
\newcommand{\shepabstract}
{We make a theoretical attempt to compare the predictions from normal
  and inverted hierarchical models, within the framework of  $\mu-\tau$ symmetry. We
  consider three major  theoretical issues in a self consistent ways, viz.,
 predictions on  neutrino mass and mixing parameters, stability under RG analysis in MSSM,
  and baryogenesis through leptogenesis. We further extend    our
earlier works on 
parametrisation of neutrino mass matrices obeying $\mu-\tau$ symmetry, using only  two
parameters in addition to an overall mass scale $m_0$, to both normal
and inverted hierarchy, and the ratio of
these two parameters  fixes   the value of  solar mixing angle.  Such
parametrisation though phenomenological, gives a firm handle on
the analysis of the mass matrices  and   can also extend its
prediction to  lower values of  solar mixing
angle  in the range 
$\tan^2\theta_{12}= 0.50 - 0.35$.  All predictions are in agreement
with observed data except in one case, i.e., inverted hierarchy with
opposite CP parity in the first two mass eigenvalues (Type B) with
$m_3\neq 0$, where $\bigtriangleup m^2_{23}$ is highly dependent on
solar mixing angle, and the prediction is good for
$\tan^2\theta_{12}\leq 0.45$ only. 
We then check the  stability  of the inverted hierarchical model with
opposite CP-parities, under radiative corrections in MSSM for
 large $\tan\beta \sim 58-60$ region and  observe that the
  evolution of $\bigtriangleup m^2_{21}$ with energy scale, is highly
  dependent on the input-high scale value of solar mixing angle. Solar angles
  predicted by tri-bimaximal mixings angle and  values  lower than
  this, do not lead to the stability
  of the model at large $\tan\beta$ values. Similarly, the evolution of the atmospheric mixing angle
  with energy scale at large $\tan\beta$ values, shows sharp decrease
  with energy scale
  for the case with $m_3\neq 0$.  However, non-zero value of  $m_3$
  is essential to 
  maintain the  stability on  the evolution of   solar mass scale. We
  apply these mass matrices to estimate the baryon asymmetry of the
  Universe in a self consistent way and find that normal hierarchical
  model leads to the best result. Considering all these three pieces of
  theoretical investigations, we may conclude that normal hierarchical
  model is more favourable in nature.\\

PACS numbers: 14.60.Pq, 12.15.Ff, 13.15.+g, 13.40.Em
}
 
%88888888888888888888888888888888888888888888888888888888888888888
\begin{titlepage}
\begin{flushright}
%hep-ph/yymmnnn
%hep-ph/0111319
\end{flushright}
\begin{center}
{\large{\bf\sheptitle}}
\bigskip \\
\shepauthor
\\
\mbox{}
\\
{\it \shepaddress}
\\
\vspace{.5in}
{\bf Abstract} \bigskip \end{center}\setcounter{page}{0}
\shepabstract
\end{titlepage}
\section{Introduction}

The recent global $3\nu$ oscillation analysis[1] indicates a mild
departure from tribimaximal neutrino mixings. The decreasing trend in
solar mixing is also consistent with the prediction from the
Quark-Lepton Complementarity (QLC) relation [2,3,4] at the unification
scale where the Cabibbo angle is taken at high scale[5]. 
In the theoretical front there are several attempts to find out the
most viable models of neutrinos, and among them  the  $\mu$-$\tau$
reflection  symmetry[6-10] in neutrino mass matrix at
high scale, has attracted considerable attentions in the last few
years. Even the tri-bimaximal mixing[11,12] is a special case of this
symmetry. It is expected that this symmetry has a strong   potential to
explain the present neutrino observed data[1].

  The $\mu$-$\tau$ symmetry
 leads to the  maximal atmospheric mixing
($\theta_{23}=\pi/4$) and zero reactor angle ($\theta_{13}=0$). The
prediction on solar mixing  angle $\theta_{12}$ remains arbitrary and it is generally fixed
through a parametrisation in the mass matrix. This symmetry has the
freedom 
to fix the  solar mixing angle at lower values, even  far  below the 
 tri-bimaximal value, without
destroying the $\mu$-$\tau$ symmetry.  This is possible through the
identification of a  ratio of two parameters (referred to as flavor twister)
present in the neutrino mass matrix and  its  subsequent variation in
  input values[13]. We are
interested  to parametrise both  inverted as well as normal
hierarchical neutrino mass  models and then identify the flavor twisters
responsible for lowering the solar mixing angle[14]. It is interesting to note
that $\mu$-$\tau$ symmetry  gives  a common origin for  both
hierarchical and inverted hierarchical neutrino mass matrices in
agreement with latest data.

The $\mu$-$\tau$ symmetry in neutrino mass matrix is assumed to hold
in the charged lepton mass basis, although the charged lepton masses
are obviously not $\mu$-$\tau$ symmetric. However, such a scenario can
be realised in gauge models with different Higgs doublets generating
the up- and down-like particle masses[7,9,10,15,16]. 
A   realisation  of   $\mu$-$\tau$
symmetry  in the flavor basis  within the
framework of  SUSY SU(5) GUT,  has strengthen
the foundation  of the symmetry as a full-fledged  gauge
theory[10]. We are now interested 
to investigate  the phenomenological predictions  of the $\mu$-$\tau$
symmetry in  neutrino sector and also possible  application in leptogenesis[7,17]. In the
present work we confine to analysis without phases, keeping our eyes
on both predictions on neutrino  masses and mixings consistent with latest data.

 The paper is organised as follows. In section 2 we give a
very brief overview on latest developments on $\mu$-$\tau$ symmetry in 
neutrino mass models. In section 3 we give the parametrisation of the mass matrices
for hierarchical and inverted hierarchical models in terms of only  two
parameters and identify the flavor twisters in both cases. This will
be supplemented by detailed numerical analysis.  In section 4 we discuss
the stability under radiative corrections for large $\tan\beta$
values for inverted hierarchical model. We give a brief account on the
predictions on baryogenesis using the same neutrino mass matrices. Section 6 concludes with a summary and discussion. 

\section{Neutrino mass matrices with  $\mu$-$\tau$  symmetry}

The $\mu$-$\tau$  symmetry in the neutrino mass matrix, implies an invariance under the simultaneous
permutation of the second and third rows as well as the second and
third columns in neutrino mass matrices[6],
\begin{equation}
 m_{LL}=
 \left(\begin{array}{ccc}
 X  &  Y  &   Y\\
  Y   &  Z  &   W \\
 Y  &  W  &  Z
 \end{array}\right).
 \end{equation}
Neutrino mass matrix in eq.(1) predicts the  maximal atmospheric
mixing angle,
$\theta_{23}=\pi/4$ and  $\theta_{13}=0$. However the prediction on
solar mixing 
angle $\theta_{12}$ is arbitrary, and  it can be fixed by  the input values of the
parameters present in the mass matrix. Thus
\begin{equation}
\tan2\theta_{12}=|\frac{2\sqrt{2}Y}{(X-Z-W)}|
\end{equation}
which  depends on four input parameters
$X,Y,Z$ and $W$. This makes us difficult to choose the values of these
free parameters for a  solution consistent with
neutrino oscillation data. This point will be addressed in section 3 where the solar
angle is made dependent  only on  the ratio of two parameters, $\eta/\epsilon$. Such
parametrization of the mass matrix enables us to analyse the neutrino
mass  matrix in a  systematic and economical way[13]. The actual
values of these two new parameters will be fixed by the data on neutrino mass
squared differences. 
 
The MNS leptonic mixing matrix $U_{MNS}$ which diagonalises  $m_{LL}$
is defined by
$m_{LL}=U_{MNS}DU^{\dagger}_{MNS}$  where $D=diag.(m_1, m_2, m_3)$, and 
 \begin{equation}
 U_{MNS}=
 \left(\begin{array}{ccc}
 U_{e1} &  U_{e2}  &   U_{e3} \\
  U_{\mu 1}  &  U_{\mu 2}  &   U_{\mu 3} \\
 U_{\tau 1}  &  U_{\tau 2}  &  U_{\tau 3}
 \end{array}\right).
 \end{equation}
 From the  consideration of $\mu$-$\tau$ reflection symmetry,
 $U_{MNS}$  mixing matrix is generally  parametrised by three rotations
($\theta_{23}=\pi/4$, $\theta_{13}=0$): 
\begin{equation}
 U_{MNS}=O_{23}O_{13}O_{12}=O_{23}O_{12}=
 \left(\begin{array}{ccc}
 c_{12} &  -s_{12}  &   0 \\
  s_{12}/\sqrt{2}  &  c_{12}/\sqrt{2}  &  - 1/\sqrt{2} \\
 s_{12}/\sqrt{2}  &  c_{12}/\sqrt{2}  &  1/\sqrt{2}
 \end{array}\right)
 \end{equation}
where $c_{12}=\cos\theta_{12}$,
$s_{12}=\sin\theta_{12}$. Tri-bimaximal mixing (TBM) is a special case with
$c_{12}=\sqrt{2/3}$ and $s_{12}=\sqrt{1/3}$,
\begin{equation}
 U_{TBM}=
 \left(\begin{array}{ccc}
 \sqrt{2/3} &  -1/\sqrt{3}  &   0 \\
  1/\sqrt{6}  &  1/\sqrt{3}  &  - 1/\sqrt{2} \\
 1/\sqrt{6}  &  1/\sqrt{3}  &  1/\sqrt{2}
 \end{array}\right)
 \end{equation}
where
\begin{equation}
 O_{23}=
 \left(\begin{array}{ccc}
 0 &  0  &   0 \\
 0  &  1/\sqrt{2}  &  - 1/\sqrt{2} \\
 0  &  1/\sqrt{2}  &  1/\sqrt{2}
 \end{array}\right),
 \end{equation}
and
\begin{equation}
 O_{12}=
 \left(\begin{array}{ccc}
 \sqrt{2/3} &  -1/\sqrt{3}  &   0 \\
  1/\sqrt{3}  &  \sqrt{2/3}  &  0 \\
 0  &  0  &  1
 \end{array}\right).
 \end{equation}
For completeness we also give the three neutrino  mass eigenvalues[9]
corresponding to  the
neutrino mass  matrix in eq.(1),
\begin{equation}
-m_1=\frac{1}{2}[Z+W+X-\sqrt{8Y^2+(Z+W-X)^2}],
\end{equation}
\begin{equation}
m_2=\frac{1}{2}[Z+W+X+\sqrt{8Y^2+(Z+W-X)^2}],
\end{equation}
\begin{equation}
m_3=(Z-W).
\end{equation}
The solar mixing  angle is given by 
$\cos\theta_{12}=\sqrt{\frac{m_2+X}{m_1+m_2}}$, $\sin\theta_{12}=\sqrt{\frac{m_1-X}{m_1+m_2}}$.
If $X=0$, then we have a simple relation,
$\tan^2\theta_{12}=m_1/m_2$. 
The general form of  mass matrix in eq.(1)  can be fitted to
both normal  and  inverted hierarchical models. Without changing
the expression for the prediction  on solar mixing angle in eq.(2),
the parameters $X$, $Z$ and $W$ can be
rearranged within  the texture, giving many possible neutrino mass
models obeying $\mu$-$\tau$ symmetry. Some of these models are
suitable for normal hierarchical model and other for inverted
hierarchical model. This will be addressed in the next two subsections. 
  
\subsection{ Normal  hierarchical neutrino mass model}
Many models based on normal hierarchy [8,9] make the 1-1 term ($X$) zero
in the general neutrino mass matrix in eq.(1). This reduces one free parameter in the mass
matrix. In the present analysis this can be done from the general form
(1) by a mere rearrangement of parameters, which preserves the  solar
mixing  angle given in eq.(2). Thus the mass matrix takes the new form  
\begin{equation}
 m_{LL}=
 \left(\begin{array}{ccc}
 0  &  Y  &   Y\\
  Y   &  Z-X/2  &   W-X/2 \\
 Y  &  W-X/2  &  Z-X/2
 \end{array}\right)
 \end{equation}
which  can be simply expressed as[8,9],
\begin{equation}
 m_{LL}=
 \left(\begin{array}{ccc}
 0  &  A  &   A \\
  A   &  B  &   C  \\
 A  &  C  &  B
 \end{array}\right).
 \end{equation}
As discussed before, such mass matrix  has interesting predictions on
solar mixing  angle in term of a ratio of
first   two  neutrino mass eigenvalues[8,9,18,19],
\begin{equation}
\tan^2\theta_{12}=\frac{m_1}{m_2}.
\end{equation}
This form of neutrino mass matrix (12)  is seesaw invariant[8] in the sense
that both Dirac neutrino mass matrix $m_{LR}$ and the right-handed Majorana mass
matrix $M_R$  also  have the same form(12) of  mass matrix.  This model can be
motivated[8] within
the framework of  SO(10) GUT where both quarks and charged leptons mass matrices
have the broken $\mu$-$\tau$ symmetry due to the presence of extra
$\mu$-$\tau$ antisymmetric parts in the mass matrices arising from
{\bf 120}
Higgs scalar, while mass mass matrices belonging to three neutrinos
$(m_{LL}, m_{LR}, M_R)$ are
$\mu$-$\tau$ symmetric due to {\bf 10} and {\bf 126} Higgs scalar in
the SO(10)GUT model. In such  model there is a  correction in $U_{MNS}$ from charged
lepton sector.

Other  interesting observations for the neutrino mass matrices (1) obeying
$\mu$-$\tau$  symmetry, can also be
obtained from the  seesaw formula using the general  $\mu$-$\tau$ symmetric Dirac
neutrino mass matrix (1)  in the diagonal basis of the right-handed Majorana
mass matrix with  two degenerate  heavy mass eigenvalues $(M_1, M_2,
M_2)$ [7]. This is true only  for normal hierarchical model and
has a special application in resonant leptogenesis.

\subsection{ Inverted hierarchical neutrino mass model}
In case of inverted hierarchical model, we assume that both CP even
and CP odd in first  two mass eigenvalues, have a common mass matrix[13]. The
general form in eq.(1) with $Z\neq W$ will lead to small but  $m_3\neq
0$. For simplicity in the phenomenological  analysis, one  can
push to  $m_3=0$ condition
without changing the expression on the prediction of solar mixing
angle (2), through a simple rearrangement of parameters. The inverted
hierarchical mass matrix has the form 
\begin{equation}
 m_{LL}=
 \left(\begin{array}{ccc}
 X  &  Y  &   Y\\
  Y   &  (Z+W)/2  &   (Z+W)/2 \\
 Y  &  (Z+W)/2  &  (Z+W)/2
 \end{array}\right).
 \end{equation}
which can be rewritten as 
\begin{equation}
 m_{LL}=
 \left(\begin{array}{ccc}
 A  &  B  &   B \\
 B   &  D  &   D  \\
 B  &  D  &  D
 \end{array}\right).
 \end{equation}
This  obeys $det( m_{LL}) = 0$ condition[20] and hence $m_3=0$. The
above form (15)  can
now be expressed as [21],
\begin{equation}
 m_{LL}=
 \left(\begin{array}{ccc}
 \delta_{1}  &  1  &   1 \\
  1   &  \delta_{2}  &   \delta_{2}\\
 1  &  \delta_{2}  &  \delta_{2}
 \end{array}\right)m_{0},
 \end{equation}
where $\delta_{1,2} <1$ for inverted hierarchy with CP odd. Particular choices of the values
 of parameters (A, B, D) in (15) make the mass matrix  either CP even $(B<D)$
 or CP odd $(B>D)$  in first two mass eigenvalues $(m_1,\pm
m_2,0)$, thus  signifying a common origin for  both mass models.
 Recently Babu et al [16] have presented a new realisation of inverted
hierarchical mass matrix (16)  based on $S_3\times U(1)$ flavor symmetry where $S_3$ is the
non-Abelian group generated by permutation of three objects, while the
$U(1)$ is based for explaining the mass hierarchy of the leptons. In
this construction the $S_3$ permutation symmetry is broken down to an
Abelian $S_2$ in the neutrino sector, whereas it is broken completely in
the charged lepton sector. The $\mu$-$\tau$ symmetry is then  realised
in neutrino sector, while having non-degenerate charged leptons. The
$U(1)$ symmetry acts as leptonic $L_e-L_{\mu}-L_{\tau}$ symmetry which is
desirable for an inverted hierarchical model. The significant  deviation of
$\theta_{12}$ from $\pi/4$ comes from breaking of $S_2$ symmetry in
charged lepton sector.

It can be pointed out here that the form of mass matrix (12) with
1-1 element zero in the mass matrix,
can also be constructed for inverted hierarchy[22]. Since  $m_1$ and $m_2$ are
nearly degenerate in inverted hierarchy, this model leads to nearly
bi-maximal mixings $\tan^2\theta=m_1/m_2\sim 0.98$, requiring large  corrections
from charged lepton sector to meet the data. Such models have problems
and are not favoured by the recent data.
  
In a significant work  by Mohapatra et al [10], the realization of $\mu$-$\tau$
reflection
  symmetry in the neutrino mass matrix
in the flavor basis (i.e. the basis where charged leptons are mass
eigenstates),
 has been obtained in a  realistic full-fledged gauge model based on
 $SUSY SU(5)GUT$ where leptons and quarks are treated together. In such
 model the requirement of $\mu$-$\tau$ symmetry for neutrinos does
 not contradict with the observed fermion masses and mixings. The  neutrino
 mass matrix having $\mu$-$\tau$ symmetry, is  assumed to
arise from a triplet seesaw (type II) mechanism, which disentangles
the neutrino flavor structure from quark flavor structure.  
The deviations of $\theta_{13}$ and $\theta_{23}$ from $0$ and $\pi/4$
respectively,  come from left-handed charged leptons mixing
matrix.

%%%%%%%%%%%%%%%%%%%%%%%%%%%%%%%%%%%%%%%%%%%%%%%%%%%%%%%%%%%%%%%%%%%%%%%%%%%
\section{Neutrino mass matrices in two parameters and numerical analysis}
This section is the main part of the paper, where we are interested
 to express the general mass matrix (1) with  only  two
parameters $\eta$ and  $\epsilon$, with an  additional   mass scale $m_0$. The
expression for the prediction on solar mixing  angle (2) will now
depend only
 on the ratio of these two variables, $\eta$ and
$\epsilon$.  Such
consideration in the reduction of the number of parameters in the
texture,  gives a firm handle on the phenomenological analysis of
the mass matrices obeying $\mu$-$\tau$ symmetry. The parametrisations
presented below  are  not unique. We give such parametrisation  for
normal 
hierarchy as well as inverted hierarchy. The numerical
analysis is performed  using Mathematica.

\subsection{Parametrisation for normal Hierarchy}
We present here  two forms of parametrisation related to the  mass
matrices in eqs.(1) and (12) discussed in section 2 and 2.1, 
with  two parameters $\eta$ and
$\epsilon$ in the texture, and  a common mass scale
parameter $m_0$.

{\bf  Case (i)  with $X\neq 0$}: The general mass matrix of the form(1)
with  no zero texture,  is parametrised  by 
\begin{equation}
 m_{LL}(NH)=
 \left(\begin{array}{ccc}
 -\eta  &  -\epsilon  &   -\epsilon \\
  -\epsilon   &  1-\epsilon  &   -1  \\
 -\epsilon  &  -1  &  1-\epsilon
 \end{array}\right)m_0
 \end{equation}
This predicts an  expression for solar mixing angle, 
$$ \tan 2\theta_{12}=\frac{2\sqrt{2}}{|\eta/\epsilon-1|}$$
where the ratio $x=\frac{\eta}{\epsilon}$ is the ``flavor twister'' in this
case. The possible solutions for lowering solar angle beyond  tribimaximal solar
mixing, are given by 
$|\eta/\epsilon -1|\leq 1$ which leads to $\eta/\epsilon \leq 0$ and $\eta/\epsilon\geq2$. 
 The numerical  predictions on deviation from tri-bimaximal mixings
 for the case 
$\eta/\epsilon\leq 0$ are presented  in  Table-1.  The  predictions on
 $\bigtriangleup m^2_{23}$ and  $\bigtriangleup m^2_{21}$ are
 consistent with observed data for  a wide range of
 solar angle $tan^2\theta_{12}=(0.50 - 0.35)$.

 As an   example we cite a
 representative case for
 $\tan^2\theta_{12}=0.45$. Taking input values for  $\eta/\epsilon=-0.1595$,
 $\epsilon=0.1894$ and $m_0=0.029eV$, we get the three mass
 eigenvalues, $m_i=(0.00609,
 -0.0107, 0.05251)eV$, leading to $\bigtriangleup m^2_{21}=7.75\times
 10^{-5}eV^2$ and $\bigtriangleup m^2_{23}=2.64\times 10^{-3}eV^2$.  

{\bf Case (ii)  with $X=0$}: Following the procedure in eq.(12), the
mass matrix (17)  is now  modified as 
\begin{equation}
 m_{LL}(NH)=
 \left(\begin{array}{ccc}
 0  &  -\epsilon  &   -\epsilon \\
  -\epsilon   &  1-\epsilon  &   -1+\eta  \\
 -\epsilon  &  -1+\eta  &  1-\epsilon
 \end{array}\right)m_0
 \end{equation}
where the solar mixing  angle prediction corresponding to a  choice of
flavor twister, is same as that of  eq.(17),
except that the relation $\tan^2\theta_{12}=\frac{m_1}{m_2}$ is valid in
this case since 1-1 term  in the mass matrix is zero. Table-2 gives the
numerical 
 predictions at lower values of  solar angle.
The predictions on solar and atmospheric mass scales are consistent
with the recent experimental data. In order to have the  result for
 $\tan^2\theta_{12}=0.45$, we  take the  input values for  $\eta/\epsilon=-0.1595$,
 $\epsilon=0.17$ and $m_0=0.028eV$, and  we get $m_i=(0.00452,
 -0.01004, 0.05199)eV$ leading to $\bigtriangleup m^2_{21}=8.03\times
 10^{-5}eV^2$ and $\bigtriangleup m^2_{23}=2.60\times 10^{-3}eV^2$ respectively.  

\subsection{Parametrisation of inverted hierarchy}
We again consider two cases with $m_3\neq 0$ and $m_3=0$ respectively  for  inverted
hierarchical model, with only two parameters $\eta$ and $\epsilon$, in
addition to an overall mass scale $m_0$. As discussed earlier, we do
not expect zero texture in 1-1 element in  inverted
hierarchical model.  

{\bf  Case (i) with $m_3\neq 0$:}
A suitable parametrisation for mass matrix (1) has the form[13],
\begin{equation}
 m_{LL}(IH)=
 \left(\begin{array}{ccc}
1 -2\epsilon  &  -\epsilon  &   -\epsilon \\
  -\epsilon   &  1/2  &   1/2-\eta  \\
 -\epsilon  &  1/2-\eta  &  1/2
 \end{array}\right)m_0.
 \end{equation}
This gives the prediction of solar angle,
 $$\tan 2\theta_{12}=\frac{2\sqrt{2}}{|(2-\eta/\epsilon)|}.$$
This leads to the condition  for lowering solar mixing angle
beyond tribimaximal mixing, $|2-\eta/\epsilon|\leq 1$ which has two
possibilities[13]:  $\eta/\epsilon \leq 1$ and $\eta/\epsilon \geq 3$.  
The corresponding numerical  predictions are given in Table-3 for three types: A, B, and C. Type
A means $m_i=(m_1, m_2, m_3)$, type B means $m_i=(m_1,- m_2, m_3)$
and type C means $m_i=(m_1, m_2, -m_3)$ respectively,  which are related
to CP phases. The result shows that these types have a common mass
matrix and hence a  common origin.

 For a  demonstration,  we cite numerical results for  
 representative cases for
 $\tan^2\theta_{12}=0.45$. Type A: Taking input values for  $\eta/\epsilon=0.8405$,
 $\eta=0.00465$, we get $m_i=(0.04918,
 0.05003, 0.00023)eV$ leading to $\bigtriangleup m^2_{21}=8.39\times
 10^{-5}eV^2$ and $\bigtriangleup m^2_{23}=2.50\times 10^{-3}eV^2$.
  
Type B: With  input values   $\eta/\epsilon=0.8405$,
 $\eta=0.58715$, we get $m_i = ( - 0.05299, 
 0.05378, 0.02936 )eV$ leading to $\bigtriangleup m^2_{21}=8.38\times
 10^{-5}eV^2$ and $\bigtriangleup m^2_{23}=2.03\times 10^{-3}eV^2$.
 
Type C: For input values for  $\eta/\epsilon=3.16$,
 $\eta=-0.017$, we have  $m_i = (0.05028, 
 0.05111,  -0.00085)eV$ leading to $\bigtriangleup m^2_{21}=8.34\times
 10^{-5}eV^2$ and $\bigtriangleup m^2_{23}=2.61\times 10^{-3}eV^2$. In
 all three cases we take $m_0=0.05eV$ as input.  
 
{\bf  Case (ii)  with $m_3= 0$:}
Following eq.(14),  we express eq.(19) in the following  form of mass matrix,
\begin{equation}
 m_{LL}(IH)=
 \left(\begin{array}{ccc}
1 -2\epsilon  &  -\epsilon  &   -\epsilon \\
  -\epsilon   &  1/2-\eta/2  &   1/2-\eta/2  \\
 -\epsilon  &  1/2-\eta/2  &  1/2-\eta/2
 \end{array}\right)m_0.
 \end{equation}
The numerical predictions are given in Table-4 (for type B), Table-5 (for type
A)  and Table-6 ( for type C) for  the range of solar angle,
$\tan^2\theta_{12}=(0.5 - 0.35)$. 

We again give corresponding results  for
 $\tan^2\theta_{12}=0.45$ in three types.

 Type A: Taking input values for  $\eta/\epsilon=0.8405$,
 $\eta=0.00445$ and $m_0=0.05eV$, we get $m_i=(0.04922,
 0.05003, 0.0)eV$ leading to $\bigtriangleup m^2_{21}=8.03\times
 10^{-5}eV^2$ and $\bigtriangleup m^2_{23}=2.50\times 10^{-3}eV^2$.  

Type B: With  input values   $\eta/\epsilon=0.8405$,
 $\eta=0.58695$ and $m_0=0.048eV$, we get $m_i=(-0.05084,
 0.05163, 0.0)eV$ leading to $\bigtriangleup m^2_{21}=8.06\times
 10^{-5}eV^2$ and $\bigtriangleup m^2_{23}=2.58\times 10^{-3}eV^2$. 

Type C: For input values for  $\eta/\epsilon=3.16$,
 $\eta=-0.01645$ and $m_0=0.05eV$, we have  $m_i = (0.05027, 
 0.05107,  0.0)eV$ leading to $\bigtriangleup m^2_{21}=8.06\times
 10^{-5}eV^2$ and $\bigtriangleup m^2_{23}=2.61\times 10^{-3}eV^2$.

%%%%%%%%%%%%%%%%%%%%%%%%%%%%%%%%%%%%%%%%%%%%%%%%%%%%%%%%%%%%%%%%%%%%%%%%%%%
\subsection{Understanding  the parametrisation in  neutrino mass
  matrices}

{\bf A: Inverted hierarchy:}
In order to understand  the form of mass matrix parametrised in eq.(19), we start with
 two parts[24] of neutrino mass matrix, $m_{LL}=m^{o}_{LL}+
\Delta m_{LL}$, which can be diagonalised by tri-bimaximal mixing
matrix (5). For the inverted hierarchy  the structure of the dominant term
 $m^{o}_{LL}$ having $\mu$-$\tau$ symmetry, is given by
\begin{equation}
m^{o}_{LL}=\left(\begin{array}{ccc}
1 & 0 & 0 \\
0 & \frac{1}{2} & \frac{1}{2} \\
0 & \frac{1}{2} & \frac{1}{2}
\end{array}\right)m_0.
\end{equation}
which is diagonalised as
\begin{equation}
O^{T}_{23}m^{o}_{LL}O_{23}=\left(\begin{array}{ccc}
1 & 0 & 0 \\
0 & 1 & 0 \\
0 & 0 & 0
\end{array}\right)m_0
\end{equation}
The second perturbative term $\Delta m_{LL}$ can also be diagonalised
by  $(O_{23}O_{12})$,
\begin{equation}
\Delta m_{LL}=\left(\begin{array}{ccc}
2 & 1 & 1 \\
1 & 0 & 1 \\
1 & 1 & 0
\end{array}\right)m_0(-\eta)
\end{equation}
where $\eta$ is a very small  parameter.
The diagonalisation with tri-bimaximal mixing matrix (5),
 
$$(O_{23}O_{12})^{T}\Delta m_{LL}
(O_{23}O_{12})=O^{T}_{12}(O^{T}_{23}\Delta m_{LL} O_{23})O_{12}$$
gives  
\begin{equation}
O^{T}_{12}O^{T}_{23}\left(\begin{array}{ccc}
2 & 1 & 1 \\
1 & 0 & 1 \\
1 & 1 & 0
\end{array}\right)O_{23}O_{12} m_0(-\eta)
\end{equation}
 \begin{equation}
=O^{T}_{12}\left(\begin{array}{ccc}
2 & \sqrt{2} & 0 \\
\sqrt{2} & 1 & 0 \\
0 & 0 & -1
\end{array}\right)O_{12}m_0(-\eta)
\end{equation}
\begin{equation}
=\left(\begin{array}{ccc}
3 & 0 & 0 \\
0 & 0 & 0 \\
0 & 0 & -1
\end{array}\right)m_0(-\eta).
\end{equation}
Thus, from eqs.(22) and (26), the diagonalisation of the total mass
matrix, 
$$U^{T}_{TBM}m_{LL}U_{TBM}= O^{T}_{23}m^{o}_{LL}O_{23}+ (O_{23}O_{12})^{T}\Delta m_{LL}
(O_{23}O_{12})$$
 leads to 
\begin{equation}
=\left(\begin{array}{ccc}
1-3\eta & 0 & 0 \\
0 & 1 & 0 \\
0 & 0 & \eta
\end{array}\right)m_0.
\end{equation}
The deviation of solar angle from tri-bimaximal mixings can be
introduced 
through the replacement $\Delta m_{LL}$ by $\Delta {m'}_{LL}$
using a flavor twister $x=\frac{ \epsilon}{\eta}$ where 

\begin{equation}
\Delta {m'}_{LL}=\left(\begin{array}{ccc}
2x & x & x \\
x & 0 & 1 \\
x & 1 & 0
\end{array}\right)m_0(-\eta)
\end{equation}
which still has $\mu$-$\tau$ symmetry. This can be diagonalised by $O_{23}$ but
$O_{12}$  is now replaced by a  new matrix $O'_{12}$. Thus  
$O'^{T}_{12}(O^{T}_{23}\Delta m'_{LL} O_{23})O'_{12}$ leads to 
\begin{equation}
O'^{T}_{12}O^{T}_{23}\left(\begin{array}{ccc}
2x & x & x \\
x & 0 & 1 \\
x & 1 & 0
\end{array}\right)O_{23}O'_{12} m_0(-\eta)
\end{equation}
 \begin{equation}
=O'^{T}_{12}\left(\begin{array}{ccc}
2x & 2x/\sqrt{2} & 0 \\
2x/\sqrt{2} & 1 & 0 \\
0 & 0 & -1
\end{array}\right)O'_{12}m_0(-\eta)
\end{equation}
\begin{equation}
=\left(\begin{array}{ccc}
(1+2x+y)/2 & 0 & 0 \\
0 & (1+2x-y)/2 & 0 \\
0 & 0 & -1
\end{array}\right)m_0(-\eta)
\end{equation}
where $y=\sqrt{1-4x+12x^2}$. The new solar angle calculated from
$O'_{12}$, is now given by 
$$\tan2\theta_{12}=|\frac{2\sqrt{2}}{2-1/x}|$$
The corresponding new mass eigenvalues for $m'_{LL}$ are calculated as 
\begin{equation}
m_{1,2}=\frac{m_{o}}{2}[2-\eta (1+2x \pm y)]; \\
m_{3}=\eta m_{o}
\end{equation}
After identification of $x=\epsilon/\eta$ , we obtain  the same  mass
matrix(19). 

{\bf B.Normal Hierarchy:}

In  case of normal hierarchy(17), we  start with  two parts[24] of neutrino mass matrix, $m_{LL}=m^{o}_{LL}+
\Delta m_{LL}$, which can be diagonalised by tribimaximal mixing
matrix(5).   The structure of the dominant term
 $m^{o}_{LL}$ having $\mu$-$\tau$ symmetry, can be taken as 

 \begin{equation}
m^{o}_{LL}=\left(\begin{array}{ccc}
0 & 0 & 0 \\
0 & 1 & -1 \\
0 & -1 & 1
\end{array}\right)m_0
\end{equation}
which can be diagonalised by 
\begin{equation}
O^{T}_{23}m^{o}_{LL}O_{23}=\left(\begin{array}{ccc}
0 & 0 & 0 \\
0 & 0 & 0 \\
0 & 0 & 2
\end{array}\right)m_0.
\end{equation}
However, the second term $\Delta m_{LL}$  which can
be diagolanised by $(O_{23}O_{12})$, can be taken as  

\begin{equation}
\Delta m_{LL}=\left(\begin{array}{ccc}
2 & 1 & 1 \\
1 & 1 & 0 \\
1 & 0 & 1
\end{array}\right)m_0(-\epsilon)
\end{equation}
where $\epsilon$ is a very small real  parameter.
The diagonalisation of eq.(35) with tribimaximal mixing matrix,  
$$(O_{23}O_{12})^{T}\Delta m_{LL}
(O_{23}O_{12})=O^{T}_{12}(O^{T}_{23}\Delta m_{LL} O_{23})O_{12}$$
leads to  
\begin{equation}
O^{T}_{12}O^{T}_{23}\left(\begin{array}{ccc}
2 & 1 & 1 \\
1 & 1 & 0 \\
1 & 0 & 1
\end{array}\right)O_{23}O_{12} m_0(-\epsilon)
\end{equation}
 \begin{equation}
=O^{T}_{12}\left(\begin{array}{ccc}
2 & \sqrt{2} & 0 \\
\sqrt{2} & 1 & 0 \\
0 & 0 & 1
\end{array}\right)O_{12}m_0(-\epsilon)
\end{equation}
\begin{equation}
=\left(\begin{array}{ccc}
3 & 0 & 0 \\
0 & 0 & 0 \\
0 & 0 & 1
\end{array}\right)m_0(-\epsilon).
\end{equation}
From eqs.(34) and (38),  the diagonalisation,
 $$U^{T}_{TBM}m_{LL}U_{TBM}= O^{T}_{23}m^{o}_{LL}O_{23}+ (O_{23}O_{12})^{T}\Delta m_{LL}
(O_{23}O_{12}),$$
 leads to 

\begin{equation}
=\left(\begin{array}{ccc}
-3\epsilon & 0 & 0 \\
0 & 0 & 0 \\
0 & 0 & (2-\epsilon)
\end{array}\right)m_0.
\end{equation}
The deviation of solar angle from tribimaximal mixings can be done
through the replacement $\Delta m_{LL}$ by $\Delta {m'}_{LL}$
using a flavor twister $x=\frac{ \eta}{2\epsilon}$,
\begin{equation}
\Delta {m'}_{LL}=\left(\begin{array}{ccc}
2x & 1 & 1 \\
1 & 1 & 0 \\
1 & 0 & 1
\end{array}\right)m_0(-\epsilon).
\end{equation}
which still obeys $\mu$-$\tau$  symmetry and can be diagonalised by $O_{23}$. Thus 
$$O'^{T}_{12}(O^{T}_{23}\Delta m'_{LL} O_{23})O'_{12}$$
 leads to 
\begin{equation}
O'^{T}_{12}O^{T}_{23}\left(\begin{array}{ccc}
2x & 1 & 1 \\
1 & 1 & 0 \\
1 & 0 & 1
\end{array}\right)O_{23}O'_{12} m_0(-\epsilon)
\end{equation}
 \begin{equation}
=O'^{T}_{12}\left(\begin{array}{ccc}
2x & \sqrt{2} & 0 \\
\sqrt{2} & 1 & 0 \\
0 & 0 & 1
\end{array}\right)O'_{12}m_0(-\epsilon)
\end{equation}
\begin{equation}
=\left(\begin{array}{ccc}
(1+2x-y)/2 & 0 & 0 \\
0 & (1+2x+y)/2 & 0 \\
0 & 0 & 1
\end{array}\right)m_0(-\epsilon)
\end{equation}
where $y=\sqrt{9-4x+4x^2}$ and new $O'_{12}$ can be obtained in
principle. The new solar mixing angle is given by 

$$\tan2\theta_{12}=\frac{2\sqrt{2}}{2x-1}$$

The corresponding new mass eigenvalues for $m'_{LL}$ are 
\begin{equation}
m_{1,2}=\frac{m_{o}}{2}[-\epsilon \pm \eta \pm \epsilon y]; \\
m_{3}=(2-\epsilon) m_{o}.
\end{equation}
After substitution of $x=\eta/(2\epsilon)$,  we
recover the earlier mass matrix
in eq.(17).  For the value $x=0$, we have the tribimaximal condition $O_{12}=O'_{12}$ leading
to $\tan^2\theta_{12}=0.5$. 

%%%%%%%%%%%%%%%%%%%%%%%%%%%%%%%%%%%%%%%%%%%%%%%%%%%%%%%%%%%%%%%%%%%%%%%%%%%%%%%%

\begin{table}[tbp]
\begin{tabular}{lllll}\\ \hline
 $\tan^2\theta_{12}$ & $\eta/\epsilon$ & range of $\epsilon$ & 
      $\bigtriangleup m^2_{21}(10^{-5}eV^2)$ & $\bigtriangleup m^2_{23}(10^{-3}eV^2)$ \\ \hline
 $0.500$ & $0.0$  &  $0.1640 - 0.1880$ & $6.77-8.83$ & $2.75-2.64$ \\
 $0.450$ & $-0.1595$  &  $0.1762 - 0.2025$ & $6.71-8.86$ & $2.70-2.59$ \\
 $0.382$ & $-0.4142$ &  $0.2080 - 0.2380$      & $6.74 - 8.82$ & $2.57-2.45$ \\ 
$0.350$ & $-0.5538$  &   $0.2356 - 0.2707$ & $ 6.72 -8.87$ & $2.46-2.31$ \\ \hline
\end{tabular}
\hfil
\caption{\footnotesize  Normal hierarchy with non-zero 1-1 term
  $(X\neq 0)$
  in the texture. Input value of $m_0=0.029eV$.  }   
\end{table}
%%%%%%%%%%%%%%%%%%%%%%%%%%%%%%%%%%%%%%%%%%%%%%%%%%%%%%%%%%%%%%%%%%%%%%%%%%%%%%%%%%%
\begin{table}[tbp]
\begin{tabular}{lllll}\\ \hline
 $\tan^2\theta_{12}$ & $\eta/\epsilon$ & range of $\epsilon$ & 
      $\bigtriangleup m^2_{21}(10^{-5}eV^2)$ & $\bigtriangleup m^2_{23}(10^{-3}eV^2)$ \\ \hline
 $0.50$ & $0.0$  &  $0.175 - 0.195$ & $7.20-8.94$ & $2.52-2.44$ \\
 $0.45$ & $-0.1595$  &  $0.160 - 0.180$ & $7.11-9.00$ & $2.64-2.57$ \\ 
$0.35$ & $-0.5538$  &   $0.135 - 0.150$ & $ 7.16 -8.85$ & $2.87-2.83$ \\ \hline
\end{tabular}
\hfil
\caption{\footnotesize  Normal hierarchy with zero 1-1 term $(X=0)$ in the texture
and $m_0=0.028eV$}   
\end{table}

%%%%%%%%%%%%%%%%%%%%%%%%%%%%%%%%%%%%%%%%%%%%%%%%%%%%%%%%%%%%%%%%%%%%%%%%%%%%%
\begin{table}[tbp]
\begin{tabular}{llllll}\\ \hline
$Type$ & $\tan^2\theta_{12}$ & $\eta/\epsilon$ & range of $\eta$ & 
      $\bigtriangleup m^2_{21}(10^{-5}eV^2)$ & $\bigtriangleup m^2_{23}(10^{-3}eV^2)$ \\ \hline
$A$ & $0.50$ & $1.0$  &  $0.0048-0.0064$ & $7.15-9.51$ & $2.50-2.50$ \\
$ B$ & $0.50$ & $1.0$ &  $0.6607-0.6618$      & $9.50-7.20$ & $1.41-1.41$ \\ 
$ C $ & $0.50$ & $3.0$  &   $-0.0187$ to $ -0.0142$ & $ 9.41-7.20$ & $2.63-2.60$ \\ \hline
$ A $ & $0.45$ & $0.8405$  &  $0.0040-0.0053$ & $7.29-9.54$ & $2.5-2.5$ \\
$ B $ & $0.45$ & $0.8405$ &  $0.5865-0.5878$      & $9.52-7.27$ & $2.03-2.03$ \\ 
$ C $ & $0.45$ & $3.1600$  &   $-0.0193$ to $ -0.0147$ & $ 9.47-7.24$ & $2.63-2.60$ \\ \hline
$ A $ & $0.35$ & $0.4462$  &  $0.0020-0.0026$ & $7.22-9.30$ & $2.51-2.51$ \\
$ B $ & $0.35$ & $0.4462$ &  $0.3622-0.3628$      & $9.43-7.22$ & $4.01-4.01$ \\ 
$ C $ & $0.35$ & $3.5500$  &   $-0.0206$ to $ -0.0157$ & $ 9.50-7.20$ & $2.63-2.60$ \\ \hline
\end{tabular}
\hfil
\caption{\footnotesize Inverted hierarchy with $m_3\neq 0$ for  Type A, 
  Type B, and Type C,  explained in the text. Input value of  $m_0=0.05eV$.  }   
\end{table}
 
\begin{table}[tbp]
\begin{tabular}{lllll}\\ \hline
$\tan^2\theta_{12}$ & $\eta/\epsilon$ & range of $\eta$ & 
      $\bigtriangleup m^2_{21}(10^{-5}eV^2)$ & $\bigtriangleup m^2_{23}(10^{-3}eV^2)$ \\ \hline
$0.500$ & $1.0$  &  $0.6601 - 0.6615$ & $8.99 - 7.09$ & $2.304 - 2.304$ \\ 
$0.450$ & $0.8405$  &  $0.5864 - 0.5875$ & $8.96 - 7.15$ & $2.665 - 2,666$ \\ 
$0.382$ & $0.5858$ &  $0.4495 - 0.4502$      & $8.86 - 7.14$ & $3.436 - 3.438$ \\ 
$0.350$ & $0.4462$  &   $0.3621 - 0.3627$ & $ 8.97 - 6.99$ & $3.995 - 3.999$ \\ \hline
\end{tabular}
\hfil
\caption{\footnotesize  Inverted hierarchy with Type B:
  $m_i=(m_1, - m_2, 0)$. Input value $m_0=0.048eV$.}

\end{table}  
\begin{table}[tbp]
\begin{tabular}{lllll}\\ \hline
$\tan^2\theta_{12}$ & $\eta/\epsilon$ & range of $\eta$ & 
      $\bigtriangleup m^2_{21}(10^{-5}eV^2)$ & $\bigtriangleup m^2_{23}(10^{-3}eV^2)$ \\ \hline
$0.50$ & $1.0000$  &  $0.0048 - 0.0060$ & $7.15 - 8.92$ & $2.500 - 2.500$ \\ 
$0.45$ & $0.8405$  &  $0.0039 - 0.0050$ & $7.05 - 9.02$ & $2.503 - 2.503$ \\  
$0.35$ & $0.4462$  &   $0.0020 - 0.0025$ & $ 7.20 - 8.98$ & $2.510 - 2.510$ \\ \hline
\end{tabular}
\hfil
\caption{\footnotesize   Inverted hierarchy with Type A: $m_i=(m_1,
m_2,0)$, $m_0=0.05eV$}

\end{table}  
\begin{table}[tbp]
\begin{tabular}{lllll}\\ \hline
$\tan^2\theta_{12}$ & $\eta/\epsilon$ & range of $\eta$ & 
      $\bigtriangleup m^2_{21}(10^{-5}eV^2)$ & $\bigtriangleup m^2_{23}(10^{-3}eV^2)$ \\ \hline
$0.50$ & $3.00$  &  $-0.0176$ to $-0.0142$ & $8.93 - 7.18$ & $2.610 - 2.596$ \\ 
$0.45$ & $3.16$  &  $-0.0182$ to $-0.0147$ & $8.93 - 7.20$ & $2.610 - 2.596$ \\  
$0.35$ & $3.55$  &   $-0.0195$ to $-0.0157$ & $ 8.99 - 7.22$ & $2.622 - 2.598$ \\ \hline
\end{tabular}
\hfil
\caption{\footnotesize  Inverted hierarchy with Type C: $m_i=(m_1,
m_2,0)$, $m_0=0.05eV$}

\end{table}  
\section{Effects of renormalisation group analysis in MSSM for large $\tan\beta$}
There are excellent papers[3,23, 25,26]  devoted to radiative corrections on
neutrino masses and mixings, and on Quark-Lepton Complementarity
relation. However  the problem with inverted hierarchy with opposite  CP-parities in the
first two mass eigenvalues (type B), is not yet settled completely. In
particular, it has been shown with analytic calculations  in Ref.[23] that for large
$\tan\beta =50$ the radiatively generated low-scale value of
$\bigtriangleup m^2_{21}$ has a negative sign and this contradicts the
experimental data. We are interested here to examine this conjecture
for high-scale input value of solar angle given by tribimaximal mixing
and below. The effect of non-zero value of $m_3$ on the evolution of
mixing angles as well as $\bigtriangleup m^2_{21}$ will be investigated
in greater details. We take high-scale input values of  $\theta_{13}=0$ and
$\theta_{23}=\pi/4$ in the numerical analysis.   

We start with   a very brief outline on the procedure of RG
analysis without phases, while referring for  details
to our earlier works[27,28]. The effects of quantum radiative corrections of neutrino masses and
mixings in MSSM, lead to the low-energy neutrino mass matrix,
\begin{equation}
 m_{LL}(t_0)\sim
 \left(\begin{array}{ccc}
 X(t_u)  &  Y(t_u)  &   Y(t_u)e^{-I_{\tau}}\\
  Y(t_u)   &  Z(t_u)  &   W(t_u)e^{-I_{\tau}}    \\
 Y(t_u)e^{-I_{\tau}}  &  W(t_u)e^{-I_{\tau}}  &  Z(t_u)e^{-2I_{\tau}}
 \end{array}\right)R_0
 \end{equation}
where 
$$ R_0=exp[(6/5)I_{g_1}+6I_{g_2}-6I_{top}],$$
$$I_{g_i}=\frac{1}{16\pi^2}\int^{t_u}_{t_0}g^2_i(t)dt,  i=1, 2, 3, $$

$$I_{h_f}=\frac{1}{16\pi^2}\int^{t_u}_{t_0}h^2_f(t)dt,  f=top, b,
\tau, $$
$$ t=ln(\mu/1 GeV),  t_0=ln(m_t/1GeV),  t_u=ln(M_U/1GeV).$$
The above analytic solution of the neutrino mass matrix at low-energy
scale is possible only where charged lepton mass matrix is
diagonal. We have also neglected $h^2_e$, $h^2_\mu$ compared to
$h^2_\tau$, and for large $\tan\beta\sim 55-60$, we can take $R_0\sim
1$ and  $c=e^{I_{\tau}}\sim 1.06$. Thus the low-energy mass matrix (45) has
the form, 
\begin{equation}
 m_{LL}(t_0)\sim
 \left(\begin{array}{ccc}
 X  &  Y  &   Y/c\\
  Y   &  Z  &   W/c \\
 Y/c  &  W/c  &  Z/c^2
 \end{array}\right)R_0.
 \end{equation}
In this approach  the neutrino mass matrix evolves as a whole from high scale to low
scale, and diagonalisation of the mass matrix at any particular energy
scale leads to the physical neutrino mass eigenvalues as well as
mixing angles. This approach is numerically consistent with other
approach where neutrino mass eigenvalues and  three mixings
angles evolve separately  through coupled RG equations. In MSSM we
have the following RG equations[29], 
\begin{equation}
\frac{d}{dt}m_i=\frac{1}{16\pi^2}[(-\frac{6}{5}g^2_1-g^2_2+6h^2_t)+2h^2_{\tau}U^2_{\tau
  i}]m_i,  i=1,2,3,
\end{equation}
\begin{equation}
\frac{ds_{12}}{dt}=\frac{1}{16\pi^2}h^2_{\tau}c_{12}[c_{23}s_{13}s_{12}U_{\tau
  1}A_{31}-c_{23}s_{13}c_{13}U_{\tau 2}A_{32}+U_{\tau 1}U_{\tau
  2}A_{21}],
\end{equation}
\begin{equation}
\frac{ds_{13}}{dt}=\frac{1}{16\pi^2}h^2_{\tau}c_{23}c^2_{13}[c_{12}U_{\tau
  1}A_{31}+s_{12}U_{\tau 2}A_{32}],
\end{equation}
\begin{equation}
\frac{ds_{23}}{dt}=\frac{1}{16\pi^2}h^2_{\tau}c^2_{23}[-s_{12}U_{\tau
  1}A_{31}+c_{12}U_{\tau 2}A_{32}]
\end{equation}
where $A_{ki}=\frac{m_k+m_i}{m_k-m_i}$ and $U_{fi}$ are the elements
in MNS matrix(3)  parametrised by (neglecting CP Dirac phase), 
\begin{equation}
 U_{MNS}=
 \left(\begin{array}{ccc}
 c_{13}c_{12}  &  c_{13}s_{12}  &   s_{13}\\
  -c_{23}s_{12}-c_{12}s_{13}s_{23}   & c_{12}c_{23}-s_{12}s_{13}s_{23}  &   c_{13}s_{23} \\
 s_{12}s_{23}-c_{12}s_{13}c_{23}  &  -c_{12}s_{23}-c_{23}s_{13}s_{12}  &  c_{13}c_{23}
 \end{array}\right)
 \end{equation}
where $s_{ij}=\sin\theta_{ij}$ and $c_{ij}=\cos\theta_{ij}$
respectively.

We follow the standard procedure for a complete numerical analysis of
 the RGEs for neutrino masses and
mixing angles in two consecutive steps (i) bottom-up running in the
first place where running of third family Yukawa couplings and three
gauge couplings in MSSM, are carried out from top-quark mass scale at
low energy end to high energy scale[30] . In the present analysis we
consider the high scale value as the unification scale, $M_U=1.6\times
10^{16}GeV$, with large $\tan\beta=60$ as input value. For simplicity
of the calculation we take approximately the SUSY breaking scale at
the top-quark mass scale $t_0=ln(m_t)$. We adopt the standard procedure
to get the values of gauge couplings at top-quark mass scale from
experimental data, using one-loop RGEs, assuming the existence of
one-light Higgs doublet and five quark flavors below  top-quark
scale. The values of three Yukawa couplings and three gauge couplings
are calculated at high unification scale. (ii) In the second top-down
approach, the runnings  of  three neutrino masses and three mixing angles are
carried out simultaneously with the  running of  Yukawa and gauge couplings,
from high to low scale, using the input values of Yukawa and gauge
couplings evaluated in the first stage of running[28].

The normal hierarchical model is almost stable under
radiative correction[3,4] and is of little interest. This is evident from the fact that the 1-1 term
in the mass matrix 
is almost zero. There is a mild increase in both solar  and
atmospheric mixing angles  while running from high to low scale. The mass splitting is
found to be  acceptable and we are not repeating the same  investigation here.

 The evolution of neutrino masses with
energy scale in case of inverted hierarchical model with opposite
CP-parities,
 is highly affected with the high scale input value of the
solar mixing angle. We observe that the model is not stable for input
value of solar angle below
$\theta_{12}=37^{\circ}$ at large $\tan\beta$ values. We summarise the
following points: 

In Fig.1  we show the level crossing of
the magnitudes of two mass eigenvalues $|m_1|$ and $|m_2|$  at around  $\theta_{21}=35.24^{\circ}$,
leading to   a negative value for  $\bigtriangleup m^2_{21}$ at low energy scale. This fact is
shown in Fig.2 for three different high scale input values of solar
mixing angles. For higher value $\tan^2\theta_{12}=0.8$, low energy value of
$\bigtriangleup m^2_{21}$ falls in the  positive range, but for
$\tan^2\theta_{12}=0.5$ and  $\tan^2\theta_{12}=0.4$, it falls in the negative
range due to level crossing of first two mass eigenvalues. The
sensitivity of the low energy   $\bigtriangleup m^2_{21}$  with high
scale input solar mixing angle is shown in Fig.3. In the analysis we
take high scale input values: $m_1=-0.04918eV$, $m_2=0.05eV$,
$m_3=0.03306eV$, $\sin\theta_{23}=0.70711$ and $\sin\theta_{13}=0$
respectively.

 The effect of non-zero input value of $m_3$  at high scale, is
 important for maintaining the  stability
of  $\bigtriangleup m^2_{21}$. Even for larger solar angle
$\tan\theta_{12}=0.8$, case with $m_3=0.033eV$ gives better result
than the case with $m_3=0$, and this point is demonstrated   in Fig.5. This definitely lies outside
the observed data. However, zero value of $m_3$ is useful for stability
of the evolution for atmospheric angle. Fig.4 depicts the evolution of
$\sin\theta_{23}$ (upper pair) for two values of $m_3=0$ and
$m_3=0.033eV$, respectively. In case of solar angle the effect  of $m_3$ is
negligible. 

In short, the inverted hierarchical model with opposite CP-parities,  is not so  stable under
RG running in MSSM for larger $\tan\beta$ region where the effect of  RG
is maximum in $\bigtriangleup m^2_{21}$ and $\sin\theta_{23}$. At low
values of $\tan\beta$, the RG effects are normally small.

%%%%%%%%%%%%%%%%%%%%%%%%%%%%%%%%%%%%%%%%%%%%%%%%%%%%%%%%%%%
\vbox{
\noindent
\hfil
\vbox{
\epsfxsize=10cm
\epsffile [130 380 510 735] {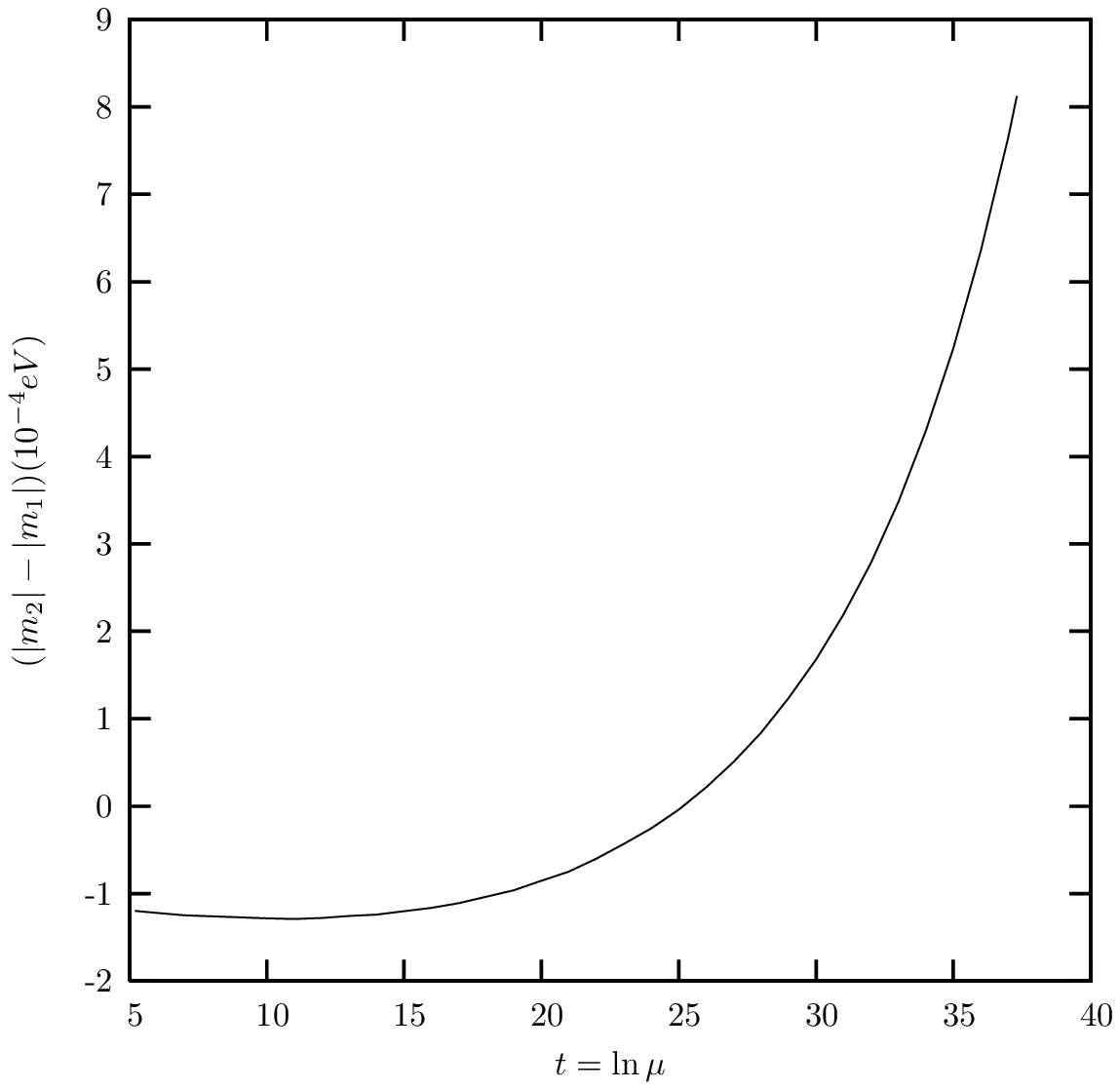}}

{\narrower\narrower\footnotesize\noindent
{Fig.1}
Evolution of the difference of the  magnitudes of  first two neutrino mass eigenvalues
$|m_1|$ and  $|m_2|$, with energy scale $\ln(\mu/1GeV)$ in inverted
hierarchy.  At lower energy scale there is the
level crossing which makes it  negative.  High scale input values are
$m_1=-0.04919 eV$,
 $m_2=0.05eV$, $m_3=0.03306 eV$, $\sin\theta_{23}=0.70711$,
 $\sin\theta_{12}=0.57735$, $\sin\theta_{13}=0$ respectively.
\par\bigskip}}

%%%%%%%%%%%%%%%%%%%%%%%%%%%%%%%%%%%%%%%%%%%%%%%%%%%%%%%%%%%
\vbox{
\noindent
\hfil
\vbox{
\epsfxsize=10cm
\epsffile [130 380 510 735] {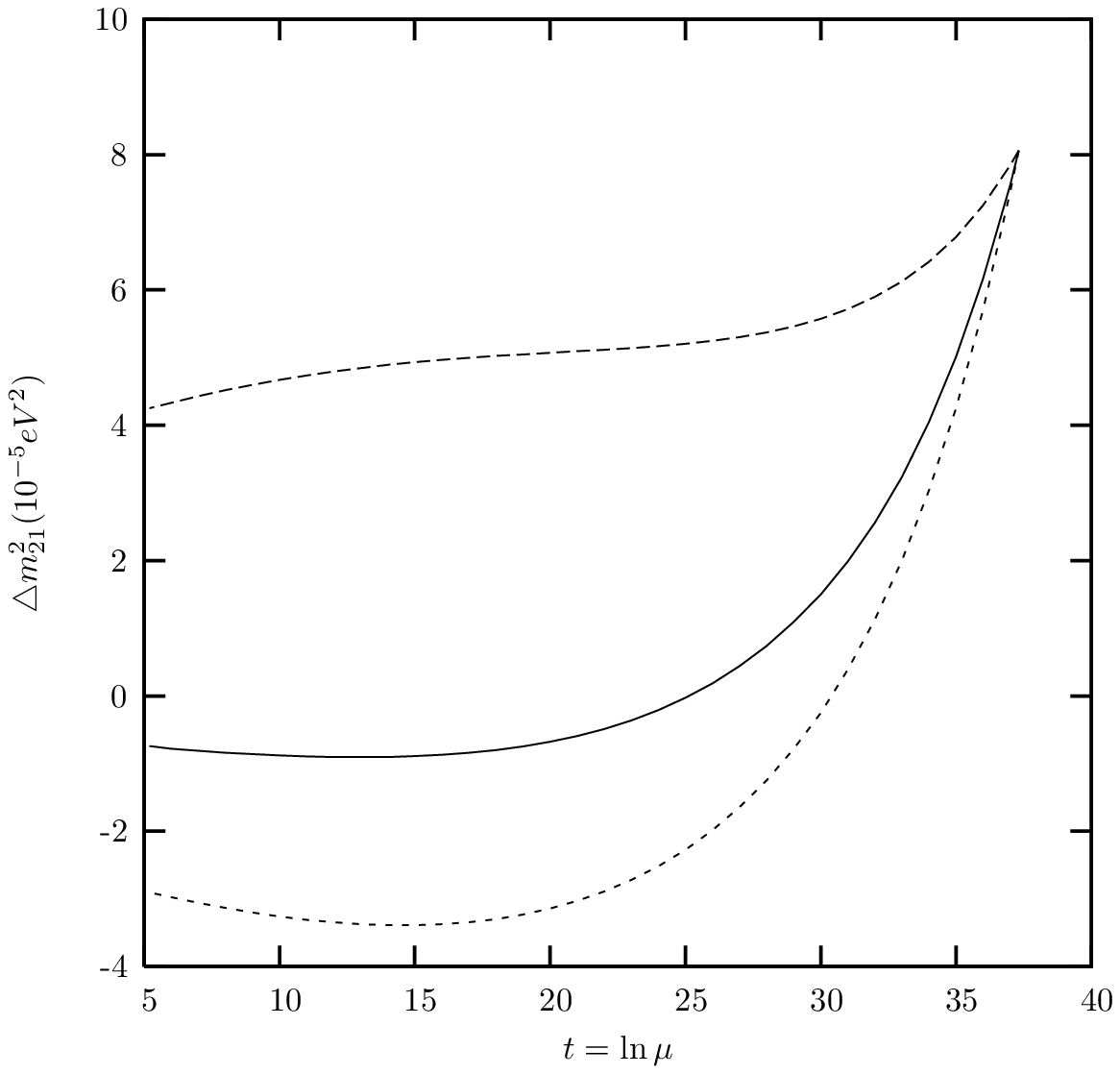}}

{\narrower\narrower\footnotesize\noindent
{Fig.2}
Evolution of $\bigtriangleup m^2_{21}$ with energy scale $\ln(\mu/1GeV)$ for different
high scale input values of  solar angle (from top to bottom): $\tan^2\theta_{12}=0.8$
(dashed-line), $\tan^2\theta_{12}=0.5$ (solid line), $\tan^2\theta_{12}=0.4$ (dotted line). 
Other input parameters are same as those  in Fig.1.  Corresponding
graphs for cases with $m_3=0$ condition,  will be more severe (see also  in Fig.5).
\par\bigskip}}
%%%%%%%%%%%%%%%%%%%%%%%%%%%%%%%%%%%%%%%%%%%%%%%%%%%%%5

%%%%%%%%%%%%%%%%%%%%%%%%%%%%%%%%%%%%%%%%%%%%%%%%%%%%%%%%%%
\vbox{
\noindent
\hfil
\vbox{
\epsfxsize=10cm
\epsffile [130 380 510 735] {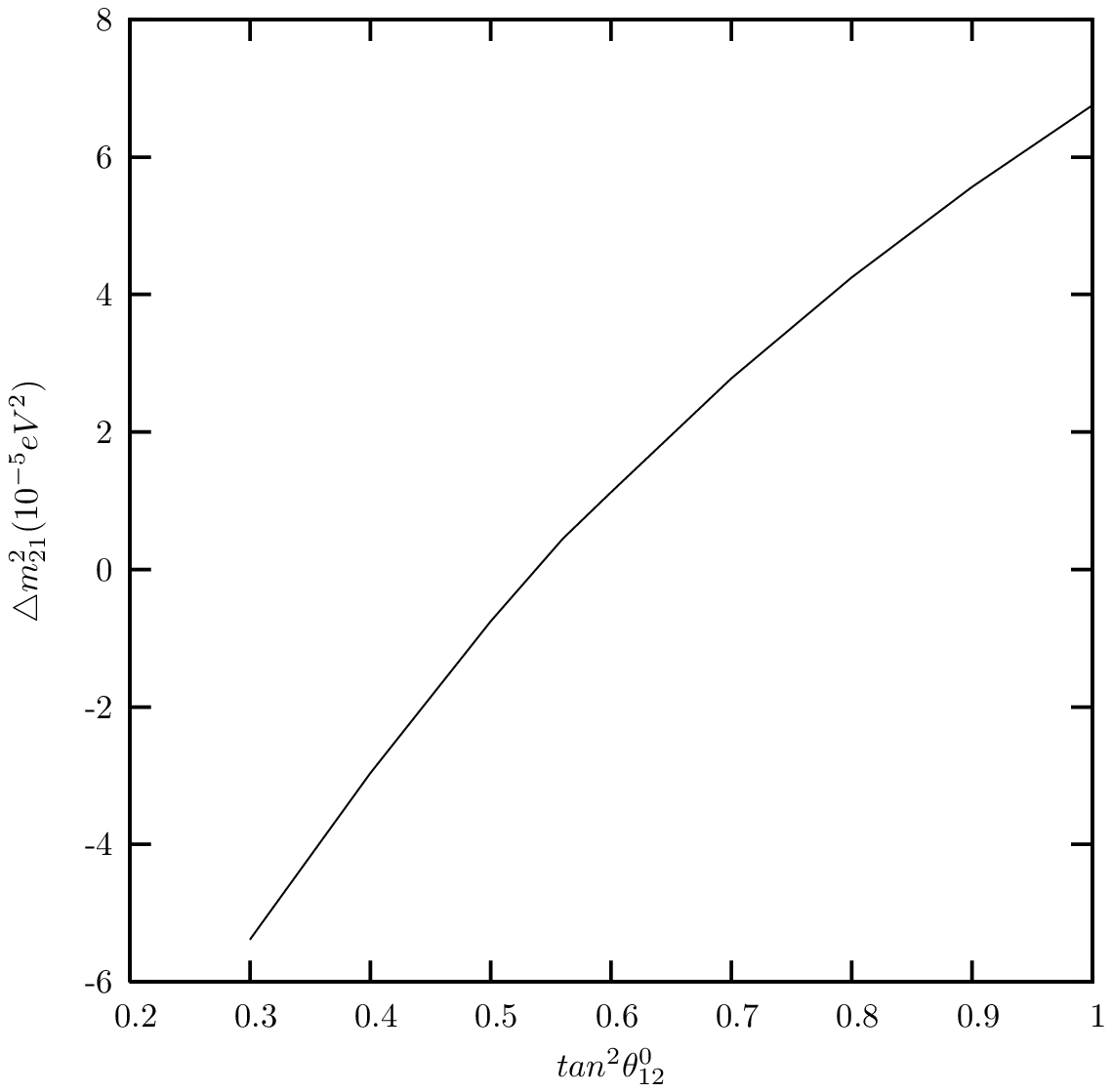}}

{\narrower\narrower\footnotesize\noindent
{Fig.3}
Sensitivity of low-energy values of $\bigtriangleup m^2_{21}$  versus  different
high scale input values of  solar mixing  angles: $\tan^2\theta_{12}$, as
analysed in Fig.2. 
Other input parameters are same as those  in Fig.1.
\par\bigskip}}
%%%%%%%%%%%%%%%%%%%%%%%%%%%%%%%%%%%%%%%%%%%%%%%%%%%%%%%%%%%%%%%%%%%%%%%%%%%%%
%%%%%%%%%%%%%%%%%%%%%%%%%%%%%%%%%%%%%%%%%%%%%%%%%%%%%%%%%%
\vbox{
\noindent
\hfil
\vbox{
\epsfxsize=10cm
\epsffile [130 380 510 735] {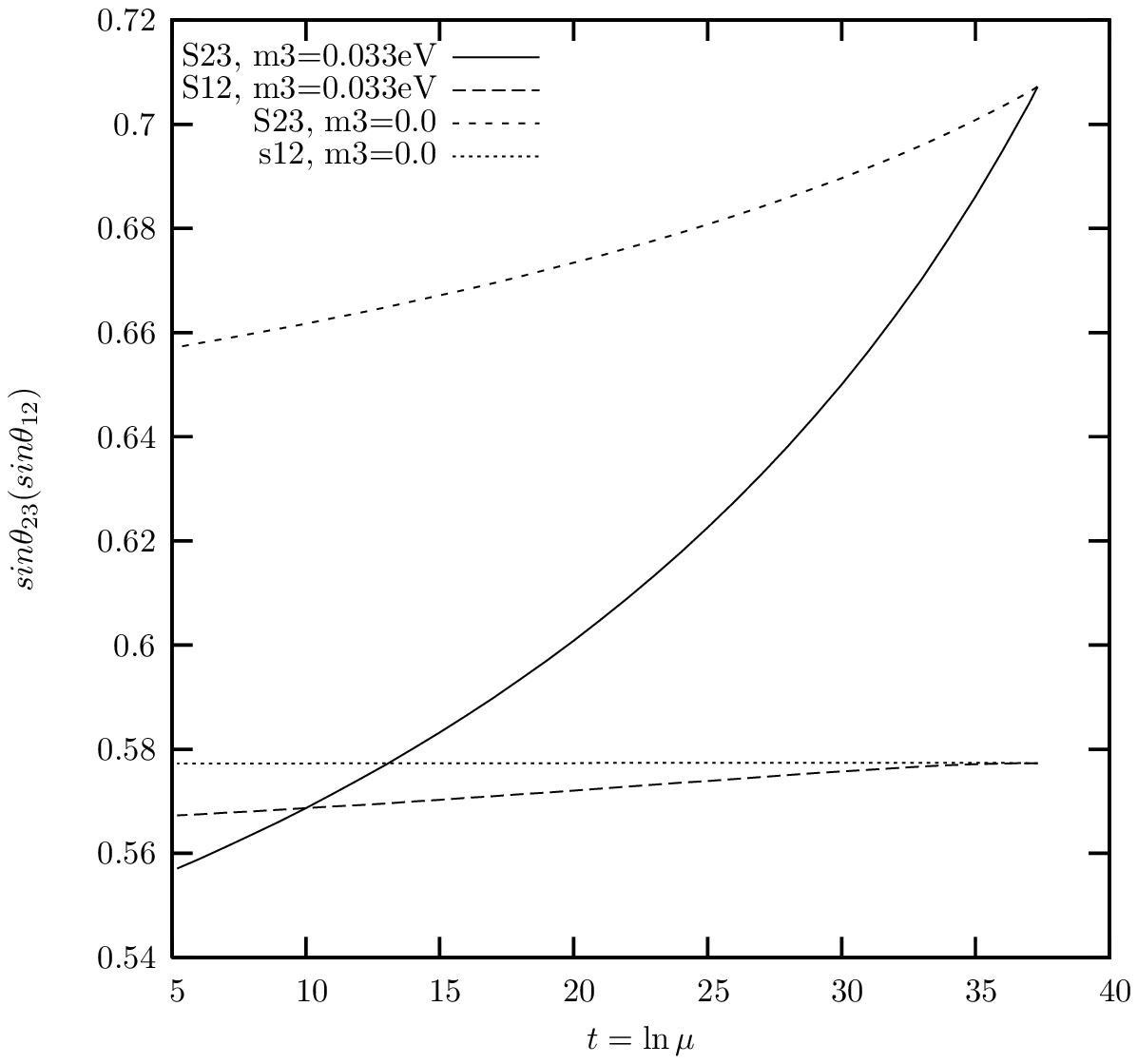}}

{\narrower\narrower\footnotesize\noindent
{Fig.4}
Evolution of $\sin\theta_{23}$  for  $m_3=0$   and 
$m_3=0.033eV$ in the first two from top; and of $\sin\theta_{12}$  for
$m_3=0$  and $m_3=0.033eV$  in the last two, respectively. Other input parameters are same as those  in Fig.1.
\par\bigskip}}
%%%%%%%%%%%%%%%%%%%%%%%%%%%%%%%%%%%%%%%%%%%%%%%%%%%%%%%%%%%%%%%%%%%%%%%%%%%%%

\vbox{
\noindent
\hfil
\vbox{
\epsfxsize=10cm
\epsffile [130 380 510 735] {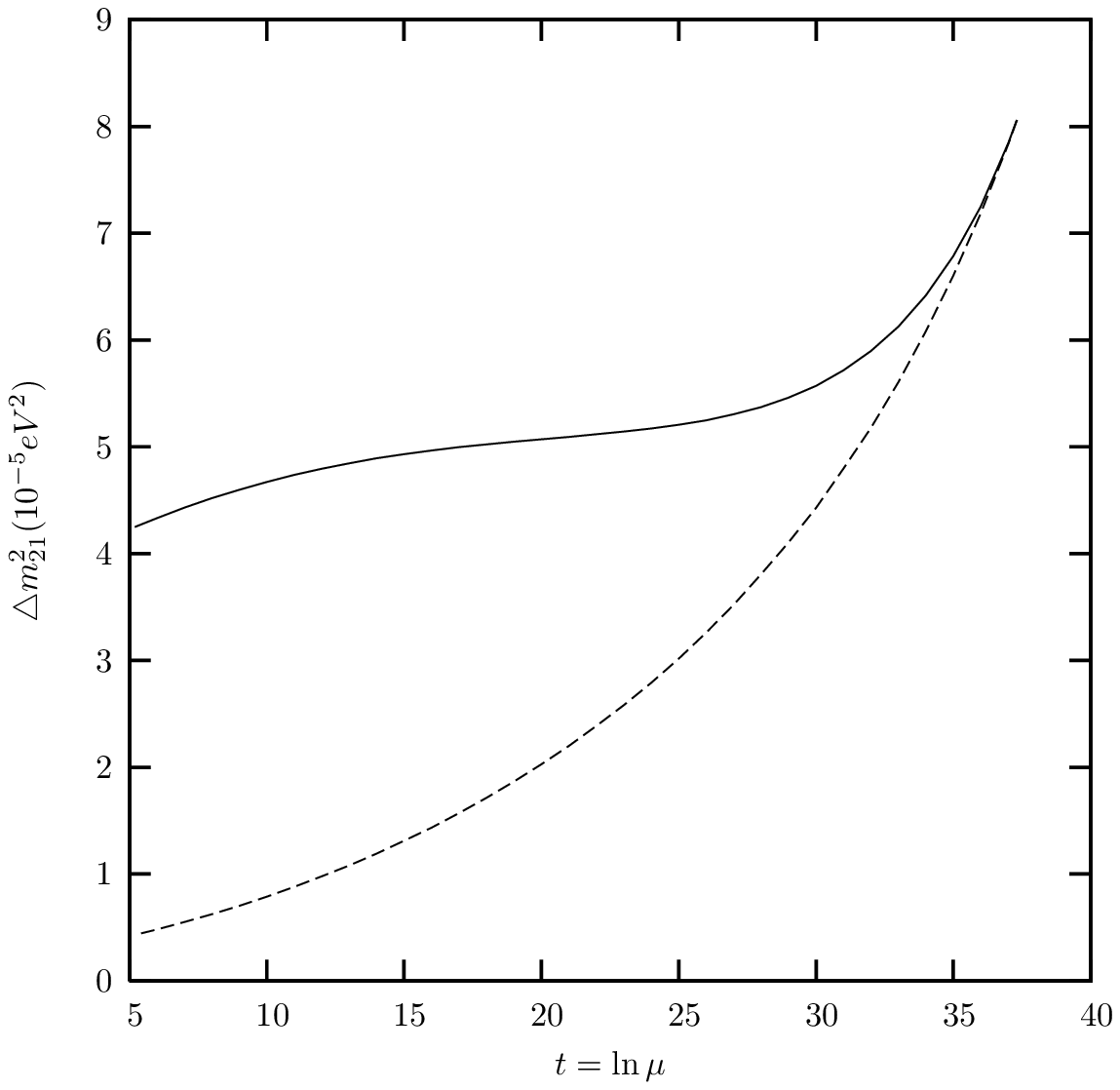}}

{\narrower\narrower\footnotesize\noindent
{Fig.5}
Evolution of $\bigtriangleup m^2_{21}$ with energy scale for two
different values of $m_3$ ( A: solid
line for 
$m_3=0.033eV$ and  B: dashed line for $m_3=0$ ) respectively. Case A
is   more stable than case B. Other high scale input values are 
same as those  in Fig.1 except for  solar mixing angle 
$\tan^2\theta_{12}=0.8$.
\par\bigskip}}
%%%%%%%%%%%%%%%%%%%%%%%%%%%%%%%%%%%%%%%%%%%%%%%%%%%%%%%%%%%%%%%%%%%%%%%%%%%%%

\section{Estimation of the baryon asymmetry of the Universe}
We present the baryon asymmetries calculated from these mass models $m_{LL}$ discussed in
section 3, using the inverse seesaw formula $M_{RR}=- m^T_{LR}m_{LL}^{-1}m_{LR}$.
 In left-right symmetric models such as SO(10) GUT, the right-handed
 neutrino in addition to its role in seesaw mechanism,
  can also explain the baryon asymmetry of the Universe through
  leptogenesis[31]. The heavy RH
 neutrino being a Majorana neutrino, can have an 
asymmetric decay into lepton and an anti-lepton with different rates
for the lepton and the anti-lepton, thereby generating 
a CP asmmetry  which is one of the Shakarov's three conditions[32] required for generating the baryon asymmetry of the Universe.
\begin{center}
$N_{R}\rightarrow l_{L}+\phi^{\dagger}$\\
$\rightarrow \overline{l}_{L}+\phi$
\end{center} 
where  $l_{L}$ is the lepton and $\bar{l}_{L}$ is the antilepton[33]. For
the temperature of the Universe less than the mass of the 
decaying lightest of the heavy RH neutrino, the out-of-equillibrium
condition is reached as the inverse decay is blocked.
 The CP asymmetry which is caused by the intereference of tree level
 with one-loop corrections for the decays of lightest 
of heavy right-handed Majorana neutrino $N_{1}$ , is defined by[33,34]
for standard model (SM) case,
\begin{equation}\label{ch902}
 \epsilon_{1} =
\frac{3}{16\pi}\left[\frac{Im[(h^{\dag}h)^{2}_{12}]}{(h^{\dag}h)_{11}}\frac{M_{1}}{M_{2}}
  +
 \frac{Im[(h^{\dag}h)^{2}_{13}]}{(h^{\dag}h)_{11}}\frac{M_{1}}{M_{3}}\right]
\end{equation}
where  $h=m^{\prime}_{LR}/v$ is the Yukawa coupling of the Dirac neutrino mass
matrix defined in the basis where the right-handed 
neutrino mass matrix is diagonal. Here $M_{1},M_{2},M_{3}$ are the
physical right-handed  Majorana masses taken in hierarchical order $(M_1<M_2<M_3)$.

For quasi-degenerate spectrum i.e., for  $M_1\simeq M_2< M_3$, the
asymmetry is largely  enhanced by a resonance factor[16,34,35]. In such situation, the lepton asymmetry is modified[34,35,36] to
\begin{equation}\label{ch306}
 \epsilon_{1} = \frac{1}{8\pi}\frac{Im[(h^{\dag}h)^{2}_{12}]}{(h^{\dag}h)_{11}} R 
\end{equation}
 where 
\begin{center}
$R=\frac{M_2^{2}(M_2^{2}-M_1^{2})}{(M_1^{2}-M_2^{2})^2 + \Gamma^2_{2} M_{1}^{2}}$ and $\Gamma_{2}=\frac{(h^{\dag}h)_{22} M_2}{8 \pi}$
\end{center}
Again, the  lepton asymmetry is further converted into baryon
asymmetry via a non-perturbative sphaleron process[37].
The ratio of baryon asymmetry to entropy $Y_B$ is related to the
lepton asymmetry through the relation, $Y_B=wY_{B-L}=\frac{w}{w-1}Y_L$
where $w=(8N_F+4N_H)/(22N_F+13N_H)$.  The baryon
asymmetry of the Universe $Y_{B}$ is defined as the ratio of 
the baryon number density $\eta_{B}$ to the photon density
$\eta_{\gamma}$ where $s=7.04 n_{\gamma}$. This can be compared with
the observational data $Y_B=(6.21\pm 0.160\times 10^{-10}$.  In SM  it can be expressed as
\begin{center}
\begin{equation}\label{ch903}
Y_{B}^{SM}\simeq d \kappa_{1} \epsilon_{1}
\end{equation}
\end{center} 
where  $ d=7.04 \times  \frac{1}{g^*_i} \frac{w}{w-1}$. For SM with $N_F=3, N_H=1,
g^*_i=106.75$, we have $d \simeq 3.62\times 10^{-2}$. This value[38] can be
compared with lower value $ d\simeq 0.98\times 10^{-2}$
 used by other authors[7,17,39].

 In the expression for baryon-to-photon ratio the efficiency factor
 (also known as dilution factor)  $\kappa_{1}$ describes
the washout factor of the lepton asymmetry  due to various lepton
number violating processes. This factor mainly depends on the effective neutrino mass $\tilde{m_{1}}$ defined by
\begin{center}
$\tilde{m_{1}}=\frac{(h^{\dag}h)_{11}v^2}{M_{1}}$
\end{center}
where v is the vev of the standard model Higgs field $v=174$ GeV.\\
For $10^{-2}eV<\tilde{m_{1}}<10^3eV$, the washout factor $\kappa_{1}$ can be well approximated by[40,41]
\begin{center}
\begin{equation}\label{ch904} 
\kappa_{1}(\tilde{m_{1}})=0.3 \left[\frac{10^{-3} eV}{\tilde{m_{1}}}\right]\left[log\frac{\tilde{m_{1}}}{10^{-3}}\right]^{-0.6}
\end{equation}
\end{center}

For numerical estimation of the baryon asymmetry we  start with the
given light Majorana neutrino mass matrix $m_{LL}$ having $\mu-\tau$
symmetry given in section $3$ and then translate this mass matrix to
$M_{RR}$ via inversion of the  seesaw formula 
$M_{RR}$=$-m_{LR}^{T}m_{LL}^{-1}m_{LR}$ where $m_{LR}$ is the Dirac
neutrino mass matrix and $M_{RR}$ is the right-handed Majorana 
neutrino mass matrix. For simplicity we consider the diagonal form of
$m_{LR}$, and as also for three different possible choices of
$m_{LR}$ namely $(i)$ up-quark mass matrix $(m,n)=(8,4)$, $(ii)$
charged-lepton mass matrix $(m,n)=(6,2)$ and $(iii)$ down-quark mass
matrix $(m,n)=(4,2)$,
 we choose a basis $U_{R}$ where $M_{RR}^{diag} = U_{R}^{T} M_{RR}
 U_{R}$=diag($M_{1},M_{2},M_{3}$) with real and positive
 eigenvalues[39].
 We transform $m_{LR}$=diag($\lambda^{m},\lambda^{n},1)v$ to the
 $U_{R}$ basis  $m_{LR} \rightarrow m'_{LR} = m_{LR} U_{R} Q$ , where 
\begin{center}
$Q = diag ( 1,e^{i \alpha}, e^{i \beta}) $
\end{center}
is the arbitrary Majorana phases responsible for CP violation.  It can
be emphasised here that the origin of  Majorana phases comes from
$M_{RR}$. Here
$v=174$ GeV is the electroweak  vev and $\lambda \simeq 0.3$ is the Wolfeinstein
paramater. For non-zero CP asymmetry the Majorana phases $(\alpha,
\beta)$ are different from $0$ and $\pi/2$. For diagonal Dirac mass
matrices, the introduction of complex  Majorana phases in MNS mixing
matrix and in the diagonalising matrix of right-handed Majorana mass
matrix,  have the same effect to get complex Dirac mass matrices
entered in the expression of CP asymmetry.
In this prime basis the Dirac neutrino Yukawa coupling becomes  $h = \frac{m'_{LR}}{v}$.
For demonstration  we first consider the normal hierarchical model as
an example given in eq.(17) for $(1,1)$ term is not zero ($X \neq 0$).

The mass eigenvalues for RH Majorana neutrinos are
$M_{RR}^{diag}=diag(3.93\times10^{11},4.09\times10^{11},2.87\times10^{14})GeV$.
 For this structure we find $Im(h^{\dagger}h)_{11}= 7.5\times10^{-3}$;
$\kappa_{1}=1.8 \times10^{-4}$. The lepton asymmetry is found out to
be $4.59\times10^{-3}$. 
Thus  following above equations (52,53) we calculate 
 $Y_{B}^{SM}$ to be $3.77\times10^{-8}$ taking $m_{LR}$ to be
 charged lepton mass matrix (case (ii)). The $m_{LL}$ has shown good
 prediction of  correct neutrino mass parameters and mixing angles
 consistent with the recent data : $\Delta m_{21}^{2}=7.1\times10^{-5}eV^{2}$,$\Delta m_{23}^{2}=2.7\times10^{-3}eV^{2}$.
The results for the other cases are given in the tables 7-16. In
the calculation we consider the three possible choices $(m,n)$ of diagonal
Dirac mass matrices $m_{LR}=diag(\lambda^m, \lambda^n,1)v$. Throughout
the calculation we fix the choice of values of the phases $(\alpha,
\beta)=(\pi/4, \pi/4)$ as it gives the maximum numerical values of
baryon asymmetry.   

In tables 8,10 and 12, the predicted values of  baryon asymmetry $Y_B$
for inverted hierarchical neutrino mass models, lie on the lower
side. Specially for Type A inverted hierarchy (IHA) we find the range  $Y_B<10^{-14}$ for
$tan^2\theta_{12}=0.50-0.35$. However Type B inverted hierarchy
(IHB) leads to slightly better result $10^{-14}<Y_B<10^{-9}$ for $tan^2\theta_{12}=0.50$ but
not so good for lower solar angles. There is still some room to
accommodate the observed data but if we consider lower value of
$d=0.96\times 10^{-2}$ then the model fails to give good results in
both inverted hierarchical models. 

From tables 14 and 16, the baryon asymmetry predictions $10^{-10}<Y_B<10^{-8}$ from normal
hierarcy (both models with $X=0$ and $X\neq 0$ cases) are consistent with observed data for wide
range of values of $tan^2\theta_{12}=0.50-0.35$. These results  can sustain
even for lower value of $d=0.96\times 10^{-2}$ parameter. 

It can be emphasised here that the present calculation of baryon
asymmetry is based on the neutrino mass matrices obeying  2-3 symmetry
for both tribimaximal mixings and deviations of solar mixing from it,
and hence only Majorana phases are considered.

%%%%%%%%%%%%%%%%%%%%%%%%%%%%%%%%%%%%%%%%%%%%%%%%%%%%%%%%%%%%%%%%%%%%%%
\begin{table}[tbp]
\begin{tabular}{llllll}\\  \hline

 $\tan^2\theta_{12}$   & type &  (m,n) & $M_{1}$&$M_{2}$&$M_{3}$  \\
\hline
 0.50 & A & (4,2)&4.01$\times$ $10^{10}$& 9.73$\times$$10^{12}$&6.25$\times$$10^{16}$\\
0.50 & A &  (6,2)&3.29$\times$ $10^{8}$&9.73$\times$$10^{12}$&6.25$\times$$10^{16}$  \\
0.50 &A &  (8,4)&2.63$\times$ $10^{6}$&7.94$\times$$10^{10}$&6.21$\times$$10^{16}$  \\
\hline

0.05 & B& (4,2)&-1.91$\times$ $10^{11}$&2.71$\times$$10^{12}$&5.59$\times$$10^{14}$   \\
0.05 & B & (6,2)&-9.99$\times$ $10^{8}$&2.63$\times$$10^{12}$&5.59$\times$$10^{14}$   \\
0.05 & B& (8,4)&-8.09$\times$ $10^{6}$&2.13$\times$$10^{10}$&5.57$\times$$10^{14}$  \\ \hline
\end{tabular}  
\hfil
\caption {\footnotesize  Inverted hierarchical models (types A,B) with
  $m_3\neq 0$, with the  predicted neutrino mass parameters,  $\bigtriangleup
m^2_{21}=7.29$$ \times 10^{-5}eV^2$ ( 8.50$\times 10^{-5}eV^2$) and  $\bigtriangleup
m^2_{23}=2.49$$\times 10^{-3}eV^2$ ( 2.30$\times 10^{-3}eV^2$) for type
 A(B) respectively. The prediction of physical right-handed Majorana
 masses in GeV for different choices of Dirac neutrino mass matrices $(m,n)$.}
\end{table}

%%%%%%%%%%%%%%%%%%%%%%%%%%%%%%%%%%%%%%%%%%%%%%%%%%%%%%%%%%%%%%%%%%%%%%%%%%%%%%%%%%%%%%%%%%%%%%%%
\begin{table}[tbp]
\begin{tabular}{lllllll}\\  \hline 

$\tan^2\theta_{12}$ &type& (m,n) & ${(h^{\dag}h)}_{11}$ & $k_{1}$ &$\epsilon_{1}$ &$Y_{B}$ \\
\hline
0.50&A&(4,2)&6.65$\times$ $10^{-5}$& 2.95$\times$$10^{-3}$&1.92$\times$$10^{-10}$&2.50$\times$$10^{-14}$\\
0.50&A&(6,2)&5.31$\times$ $10^{-7}$& 2.95$\times$$10^{-3}$&1.56$\times$$10^{-12}$&2.02$\times$$10^{-16}$\\
0.50&A&(8,4)&4.30$\times$ $10^{-9}$& 2.95$\times$$10^{-3}$&1.26$\times$$10^{-14}$&1.64$\times$$10^{-18}$\\
\hline
0.50&B&(4,2)&5.62$\times$ $10^{-4}$& 8.83$\times$$10^{-4}$&3.05$\times$$10^{-5}$&1.18$\times$$10^{-9}$\\
0.50&B&(6,2)&5.01$\times$ $10^{-6}$& 8.83$\times$$10^{-4}$&2.69$\times$$10^{-7}$&1.05$\times$$10^{-11}$\\
0.50&B&(8,4)&4.06$\times$ $10^{-8}$& 8.83$\times$$10^{-4}$&2.18$\times$$10^{-9}$&8.42$\times$$10^{-14}$\\ \hline
\end{tabular}
\hfil
\caption {\footnotesize Predictions of  $\epsilon_{1}$ the
CP asymmetry, 
$Y_{B}$ the baryon asymmetry for inverted hierarchical models (A,B) in
Table-7, with
$m_{3}\neq $ 0, for $\tan^{2}\theta_{12}$ =0.50.}
\end{table}

\begin{table}[tbp]
\begin{tabular}{llllll}\\   \hline

$\tan^2\theta_{12}$ &type& (m,n) & $M_{1}$&$M_{2}$&$M_{3}$ \\
\hline
0.45&A&(4,2)&4.01$\times$ $10^{10}$& 9.73$\times$$10^{12}$&6.59$\times$$10^{16}$\\
0.45&A&(6,2)&3.25$\times$ $10^{8}$& 9.73$\times$$10^{12}$&6.59$\times$$10^{16}$\\
0.50&A&(8,4)&2.63$\times$ $10^{6}$& 7.94$\times$$10^{10}$&6.54$\times$$10^{16}$\\
\hline
0.45&B&(4,2)&-9.76$\times$ $10^{10}$& 2.89$\times$$10^{12}$&6.23$\times$$10^{14}$\\
0.45&B&(6,2)&-8.01$\times$ $10^{8}$& 2.82$\times$$10^{12}$&6.23$\times$$10^{14}$\\
0.45&B&(8,4)&-6.56$\times$ $10^{6}$&
2.29$\times$$10^{10}$&6.21$\times$$10^{14}$\\  \hline
\end{tabular} 
\hfil
\caption{\footnotesize Inverted hierarchical models (types A,B) with
  $m_3\neq 0$ for $\tan^2\theta_{12} =0.45$ with the predicted  neutrino mass parameters, $\bigtriangleup
m^2_{21}=8.39$$ \times 10^{-5}eV^2$ ( 8.30$\times 10^{-5}eV^2$) and  $\bigtriangleup
m^2_{23}=2.51$$\times 10^{-3}eV^2$ ( 2.90$\times 10^{-3}eV^2$) for type
 A(B) respectively. The predictions of physical right-handed Majorana
 neutrino masses in GeV for different choices of Dirac neutrino mass matrices
$(m,n)$.} 
\end{table}

%%%%%%%%%%%%%%%%%%%%%%%%%%%%%%%%%%%%%%%%%%%%%%%%%%%%%%%%%%%%%%%%%%%%%%%55555

\begin{table}[tbp]
\begin{tabular}{lllllll}\\   \hline 

$\tan^2\theta_{12}$ &type& (m,n) & ${(h^{\dag}h)}_{11}$ & $k_{1}$ &$\epsilon_{1}$ &$Y_{B}$ \\
\hline
0.45&A&(4,2)&6.56$\times$ $10^{-5}$& 2.90$\times$$10^{-3}$&2.46$\times$$10^{-10}$&3.20$\times$$10^{-14}$\\
0.45&A&(6,2)&5.31$\times$ $10^{-7}$& 2.90$\times$$10^{-3}$&2.04$\times$$10^{-12}$&2.59$\times$$10^{-16}$\\
0.45&A&(8,4)&4.30$\times$ $10^{-9}$& 2.90$\times$$10^{-3}$&1.61$\times$$10^{-14}$&2.09$\times$$10^{-18}$\\
\hline
0.45&B&(4,2)&4.38$\times$ $10^{-4}$& 8.83$\times$$10^{-4}$&2.10$\times$$10^{-5}$&8.66$\times$$10^{-10}$\\
0.45&B&(6,2)&3.82$\times$ $10^{-6}$& 8.83$\times$$10^{-4}$&1.97$\times$$10^{-7}$&7.64$\times$$10^{-12}$\\
0.45&B&(8,4)&3.09$\times$ $10^{-8}$& 8.83$\times$$10^{-4}$&1.59$\times$$10^{-9}$&6.18$\times$$10^{-14}$\\ \hline
\end{tabular}
\hfil
\caption{\footnotesize  Predictions of  $\epsilon_{1}$ the CP asymmetry, $Y_{B}$ the baryon
asymmetry for 
inverted hierarchical modelS (A,B) presented in Table-9 with $m_{3}\neq  0$ for $\tan^2\theta_{12}
=0.45$}
\end{table}

%%%%%%%%%%%%%%%%%%%%%%%%%%%%%%%%%%%%%%%%%%%%%%%%%%%%%%%%%%%%%%%%%%%%%%%%%%%%%%%%%%%%%%%%%%%%%%%%%%%%%%%%%%%%%%%%

\begin{table}[tbp]
\begin{tabular}{llllll} \\ \hline

$\tan^2\theta_{12}$ &type& (m,n) & $M_{1}$&$M_{2}$&$M_{3}$ \\
\hline
0.35&A&(4,2)&4.01$\times$ $10^{10}$& 9.73$\times$$10^{12}$&1.39$\times$$10^{17}$\\
0.35&A&(6,2)&3.25$\times$ $10^{8}$& 9.73$\times$$10^{12}$&1.39$\times$$10^{17}$\\
0.35&A&(8,4)&2.63$\times$ $10^{6}$& 7.94$\times$$10^{10}$&1.37$\times$$10^{17}$\\
\hline
0.35&B&(4,2)&-6.27$\times$ $10^{10}$& 1.37$\times$$10^{12}$&9.50$\times$$10^{14}$\\
0.35&B&(6,2)&-5.15$\times$ $10^{8}$& 3.14$\times$$10^{12}$&9.50$\times$$10^{14}$\\
0.35&B&(8,4)&-4.17$\times$ $10^{6}$&
2.55$\times$$10^{10}$&9.45$\times$$10^{14}$\\ \hline
\end{tabular} 
\hfil
\caption{\footnotesize   Inverted hierarchical models (types A,B) with
  $m_3\neq 0$ for $\tan^2\theta_{12} =0.35$ with the predicted
  neutrino mass parameters, $\bigtriangleup
m^2_{21}=7.90$$ \times 10^{-5}eV^2$ ( 8.65$\times 10^{-5}eV^2$) and  $\bigtriangleup
m^2_{23}=2.50$$\times 10^{-3}eV^2$ ( 4.0 $\times 10^{-3}eV^2$) for type
 A(B) respectively. The predictions of physical right-handed Majorana
 masses in GeV for different choices of Dirac mass matrices $(m,n)$.}    
\end{table}

\begin{table}[tbp]
\begin{tabular}{lllllll}\\ \hline

$\tan^2\theta_{12}$ &type& (m,n) & ${(h^{\dag}h)}_{11}$ & $k_{1}$ &$\epsilon_{1}$ &$Y_{B}$ \\
\hline
0.35&A&(4,2)&6.56$\times$ $10^{-5}$& 2.90$\times$$10^{-3}$&4.20$\times$$10^{-10}$&2.54$\times$$10^{-14}$\\
0.35&A&(6,2)&5.31$\times$ $10^{-7}$& 2.90$\times$$10^{-3}$&1.58$\times$$10^{-12}$&2.05$\times$$10^{-16}$\\
0.35&A&(8,4)&4.30$\times$ $10^{-9}$& 2.90$\times$$10^{-3}$&2.82$\times$$10^{-14}$&1.66$\times$$10^{-18}$\\
\hline
0.35&B&(4,2)&2.76$\times$ $10^{-4}$& 8.83$\times$$10^{-4}$&1.26$\times$$10^{-5}$&4.88$\times$$10^{-10}$\\
0.35&B&(6,2)&2.32$\times$ $10^{-6}$& 8.83$\times$$10^{-4}$&1.08$\times$$10^{-7}$&4.19$\times$$10^{-12}$\\
0.35&B&(8,4)&1.88$\times$ $10^{-8}$& 8.83$\times$$10^{-4}$&8.73$\times$$10^{-10}$&3.39$\times$$10^{-14}$\\ \hline
\end{tabular} 
\hfil
\caption{\footnotesize  Predictions of  $\epsilon_{1}$ the CP asymmetry, $Y_{B}$ the baryon
asymmetry for 
inverted hierarchical models (A,B) presented in Table-11 with $m_{3}\neq  0$ for $\tan^2\theta_{12}
=0.35$.}
\end{table}
%%%%%%%%%%%%%%%%%%%%%%%%%%%%%%%%%%%%%%%%%%%%%%%%%%%%%%%%%%%%%%%%%%%%%%%%%%%%%%%%%%%%%%%%%%%%%%%%%%%%%%%%%%%%%%%%%%%%%%%%

\begin{table}[tbp]
\begin{tabular}{ccccc} \\ \hline
$\tan^{2}\theta_{12}$&(m,n)&$M_{1}$&$M_{2}$&$M_{3}$\\
\hline
0.50 & (4,2) &3.59$\times$ $10^{12}$&-5.48$\times$$10^{12}$&2.89$\times$ $10^{14}$\\
0.50 & (6,2)& 3.93$\times$ $10^{11}$&-4.09$\times$ $10^{11}$&2.87$\times$ $10^{14}$\\
0.50 & (8,4) &3.19$\times$ $10^{9}$&-3.22$\times$ $10^{9}$&2.85$\times$ $10^{14}$\\
\hline
0.45 & (4,2) &1.83$\times$ $10^{12}$&-2.64$\times$ $10^{13}$&9.79$\times$ $10^{13}$\\
0.45 & (6,2) &1.93$\times$ $10^{10}$&-2.11$\times$ $10^{13}$&9.43$\times$ $10^{13}$\\
0.45 & (8,4) &1.56$\times$ $10^{8}$&-2.21$\times$ $10^{11}$&7.28$\times$ $10^{13}$\\
\hline
0.35 & (4,2) &4.81$\times$ $10^{11}$&2.48$\times$ $10^{13}$&-1.80$\times$ $10^{14}$\\
0.35 & (6,2) &4.01$\times$ $10^{9}$&2.43$\times$ $10^{13}$&-1.80$\times$ $10^{14}$\\
0.35 & (8,4) &3.24$\times$ $10^{7}$&2.29$\times$
$10^{14}$&-1.55$\times$ $10^{14}$\\ \hline
\end{tabular}
\hfil
\caption{\footnotesize   Predictions of 
  $M_{1},M_{2},M_{3}$ in GeV  for different $(m,n)$ for Dirac mass matrices, in the normal
hierarchical model for case $X=m(1,1)\neq 0$.  The predicted  neutrino mass
parameters are  $\bigtriangleup m^2_{21}(\bigtriangleup m^2_{23})=
7.1\times 10^{-5}eV^2 (2.1\times 10^{-3}eV^2), 6.68\times
10^{-5}eV^2(2.67\times 10^{-3}eV^2),
6.9\times 10^{-5}eV^2(2.4\times 10^{-3}eV^2$ for $tan^2\theta_{12}=0.50,0.45,0.35$
respectively.}   
\end{table}
%%%%%%%%%%%%%%%%%%%%%%%%%%%%%%%%%%%%%%%%%%%%%%%%%%%%%%%%%%%%%%%%%%%%%%%%%%%%%%%%%%%%%%%%%%%%%%%%%%%%%%%%%%%%%%%%%%%%%%%%%%%%%%%%%%%%%%
\begin{table}[tbp]
\begin{tabular}{ccccccc}\\ \hline
$\tan^{2}\theta_{12}$&(m,n)&$\tilde{m}$& ${(h^{\dag}h)}_{11}$&$\kappa_{1}$&$\epsilon_{1}$&$Y_{B}$\\
\hline
0.50&(4,2)&4.3$\times$ $10^{-11}$&5.1$\times$$10^{-3}$&3.4$\times$$10^{-3}$&4.50$\times$ $10^{-4}$&6.75$\times$ $10^{-8}$\\
0.50&(6,2)&5.8$\times$ $10^{-10}$&7.5$\times$$10^{-3}$&1.8$\times$$10^{-4}$&4.59$\times$ $10^{-3}$&3.77$\times$ $10^{-8}$\\
0.50&(8,4)&5.8$\times$ $10^{-10}$&6.1$\times$$10^{-5}$&1.8$\times$$10^{-4}$&3.62$\times$ $10^{-5}$&2.96$\times$ $10^{-10}$\\
\hline
0.45&(4,2)&4.1$\times$ $10^{-11}$&2.4$\times$$10^{-3}$&3.7$\times$$10^{-3}$&5.54$\times$ $10^{-4}$&9.03$\times$ $10^{-8}$\\
0.45&(6,2)&6.6$\times$ $10^{-11}$&4.2$\times$$10^{-5}$&2.0$\times$$10^{-3}$&8.63$\times$ $10^{-6}$&7.56$\times$ $10^{-10}$\\
0.45&(8,4)&6.6$\times$ $10^{-11}$&3.4$\times$$10^{-7}$&2.0$\times$$10^{-3}$&1.47$\times$ $10^{-7}$&1.28$\times$ $10^{-11}$\\
\hline
0.35&(4,2)&2.8$\times$ $10^{-11}$&4.5$\times$$10^{-4}$&5.2$\times$$10^{-3}$&8.85$\times$ $10^{-5}$&2.07$\times$ $10^{-8}$\\
0.35&(6,2)&3.0$\times$ $10^{-11}$&3.9$\times$$10^{-6}$&5.2$\times$$10^{-3}$&3.59$\times$ $10^{-7}$&7.93$\times$ $10^{-11}$\\
0.35&(8,4)&3.0$\times$ $10^{-11}$&3.2$\times$$10^{-8}$&5.2$\times$$10^{-3}$&8.99$\times$ $10^{-9}$&2.07$\times$ $10^{-12}$\\
\hline
\end{tabular}
\hfil
\caption{\footnotesize Predictions of  $\epsilon_{1}$ and $Y_{B}$ for various values of $\tan^{2}\theta_{12}$ for normal hierarchical model,
Case(i)X=m(1,1)$\neq$ 0 presented in Table-13 .} 
\end{table}
%%%%%%%%%%%%%%%%%%%%%%%%%%%%%%%%%%%%%%%%%%%%%%%%%%%%%%%%%%%%%%%%%%%%%%%%%%%%%%%%%%%%%%%%%%%%%%%%%%%%%%%%%%%%%%%%%%%%%%%%%%%%%%%%%%%%%%%%%%%%%%%%

\begin{table}[tbp]
\begin{tabular}{ccccc}\\ \hline
$\tan^{2}\theta_{12}$&(m,n)&$M_{1}$&$M_{2}$&$M_{3}$\\
\hline
0.50 & (4,2) &3.57$\times$ $10^{12}$&-5.29$\times$$10^{12}$&3.01$\times$ $10^{14}$\\
0.50 & (6,2)& 3.85$\times$ $10^{11}$&-3.99$\times$ $10^{11}$&2.99$\times$ $10^{14}$\\
0.50 & (8,4) &3.13$\times$ $10^{9}$&-3.25$\times$ $10^{9}$&2.97$\times$ $10^{14}$\\
\hline
0.45 & (4,2) &3.72$\times$ $10^{12}$&-5.67$\times$ $10^{12}$&2.95$\times$ $10^{14}$\\
0.45 & (6,2) &4.07$\times$ $10^{11}$&-4.23$\times$ $10^{11}$&2.93$\times$ $10^{14}$\\
0.45 & (8,4) &3.31$\times$ $10^{9}$&-3.44$\times$ $10^{9}$&2.91$\times$ $10^{14}$\\
\hline
0.35 & (4,2) &4.28$\times$ $10^{12}$&-7.25$\times$ $10^{12}$&2.84$\times$ $10^{14}$\\
0.35 & (6,2) &4.92$\times$ $10^{11}$&-5.16$\times$ $10^{11}$&2.81$\times$ $10^{14}$\\
0.35 & (8,4) &4.00$\times$ $10^{9}$&-4.21$\times$
$10^{9}$&2.79$\times$ $10^{14}$\\ \hline
\end{tabular}
\hfil
\caption{\footnotesize  Predictions of 
  $M_{1},M_{2},M_{3}$ in GeV  for different $(m,n)$ in Dirac mass matrices, in the normal
hierarchical model for case $X=m(1,1)= 0$.  The predicted neutrino mass
parameters are  $\bigtriangleup m^2_{21}(\bigtriangleup m^2_{23})=
7.5\times 10^{-5}eV^2(2.4\times 10^{-3}eV^2), 7.9\times
10^{-5}eV^2(2.59\times 10^{-3}eV^2),
7.2\times 10^{-5}eV^2(2.8\times 10^{-3}eV^2)$ for $tan^2\theta_{12}=0.50,0.45,0.35$ respectively             }
\end{table}
%%%%%%%%%%%%%%%%%%%%%%%%%%%%%%%%%%%%%%%%%%%%%%%%%%%%%%%%%%%%%%%%%%%%%%%%%%%%%%%%%%%%%%%%%%%%%%%%%%%%%%%%%%%%%%%%%%%%%%%%%%%%%%%%%%%%%%%%%%%%%%%%%%%%%%%%%%%%5
\begin{table}[tbp]
\begin{tabular}{ccccccc}\\ \hline
$\tan^{2}\theta_{12}$&(m,n)&$\tilde{m}$& ${(h^{\dag}h)}_{11}$&$\kappa_{1}$&$\epsilon_{1}$&$Y_{B}$\\
\hline
0.50&(4,2)&4.4$\times$ $10^{-11}$&5.2$\times$$10^{-3}$&3.6$\times$$10^{-3}$&5.08$\times$ $10^{-4}$&8.09$\times$ $10^{-8}$\\
0.50&(6,2)&5.9$\times$ $10^{-10}$&7.0$\times$$10^{-3}$&1.8$\times$$10^{-4}$&4.91$\times$ $10^{-3}$&3.93$\times$ $10^{-8}$\\
0.50&(8,4)&5.9$\times$ $10^{-10}$&6.2$\times$$10^{-5}$&1.8$\times$$10^{-4}$&3.88$\times$ $10^{-5}$&3.09$\times$ $10^{-10}$\\
\hline
0.45&(4,2)&4.1$\times$ $10^{-11}$&5.1$\times$$10^{-3}$&3.5$\times$$10^{-3}$&4.39$\times$ $10^{-4}$&6.83$\times$ $10^{-8}$\\
0.45&(6,2)&5.6$\times$ $10^{-10}$&7.5$\times$$10^{-3}$&1.9$\times$$10^{-4}$&4.6$\times$ $10^{-3}$&3.90$\times$ $10^{-8}$\\
0.45&(8,4)&5.6$\times$ $10^{-10}$&6.1$\times$$10^{-5}$&1.9$\times$$10^{-4}$&3.68$\times$ $10^{-5}$&3.09$\times$ $10^{-10}$\\
\hline
0.35&(4,2)&3.6$\times$ $10^{-11}$&5.2$\times$$10^{-4}$&4.2$\times$$10^{-3}$&2.2$\times$ $10^{-4}$&4.47$\times$ $10^{-8}$\\
0.35&(6,2)&4.5$\times$ $10^{-10}$&7.3$\times$$10^{-3}$&2.4$\times$$10^{-4}$&3.75$\times$ $10^{-4}$&4.03$\times$ $10^{-8}$\\
0.35&(8,4)&4.5$\times$
$10^{-10}$&6.0$\times$$10^{-5}$&2.4$\times$$10^{-4}$&2.98$\times$
$10^{-5}$&3.18$\times$ $10^{-10}$\\ \hline
\end{tabular} 
\hfil
\caption{\footnotesize   Predictions of   $\epsilon_{1}$
  and $Y_{B}$ for various values of $\tan^{2}\theta_{12}$ for normal
  hierarchical model with Case(ii)X=0 presented in Table-15 }
\end{table}

%%%%%%%%%%%%%%%%%%%%%%%%%%%%%%%%%%%%%%%%%%%%%%%%%%%%%%%%%%%%%%%%%%%%%%%%%%%%%%
\section{Summary and discussions}
We summerise the main points in this work. We give a very brief
overview on the phenomenology of the neutrino mass matrices obeying
$\mu$-$\tau$  symmetry[6]. Different neutrino mass models based
on normal as well as inverted hierarchy, are outlined. We then  introduce
the 
parametrisation of mass matrices with only  two parameters, and their
ratio  determines
 the value of the  solar mixing angle. The
actual values of these parameters are then  fixed  by the  experimental bounds
on neutrino mass scales. Such parametrisation not only gives a firm handle on
the analysis of the mass matrices  but also  lowers the  solar mixing
angle up to the range  $\tan^2\theta_{12}=0.50 - 0.35$ without
affecting atmospheric and Chooz mixing angles from the tribimaximal mixings. The detailed numerical analysis
  for different forms of mass matrices obeying the $\mu$-$\tau$
  reflection  symmetry,  are given
  in Tables 1-6. However such treatment for degenerate mass matrices
  are difficult to realise in nature. All the predictions are in
  excellent agreement with data except inverted hierarchy type B in
  table 3 where $m_3\neq 0$. Here $\bigtriangleup m^2_{23}$ is highly
  dependent on solar mixing angle, and has best predicted value at
  around $\tan^2\theta_{12}\leq 0.45$.

  We also address very briefly the stability question
  of the neutrino mass models under radiative corrections in MSSM, particularly the inverted hierarchy
  with CP odd in the first two mass eigenvalues. We find that for large
  $\tan\beta\sim 58-60$, the model is not stable under RG running. 
 The evolution of solar mass scale with energy is highly
  dependent on the input high scale value of solar angle. Solar angle
  predicted by tribimaximal angle and below, does not lead to stability
  of the model. Similarly, the evolution of atmospheric mixing angle
  with energy scale  shows sharp  decrease for the case $m_3\neq 0$
  condition, making  the model unstable. However non-zero value of $m_3$
  maintains  the  stability of the  evolution of  solar mass
  scale. this implies that tribimaximal mixings with inverted
  hierarchy are not so stable under RG analysis in MSSM. Normal
  hierarchical models are generally stable under RG analysis in MSSM
  whereas inverted hierarchy type A models are highly unstable.  In a self
  consistent way we apply these mass matrices for the  prediction of  baryon
  asymmetry of the universe via leptogenesis and we find that only normal hierarchical model
  gives good results consistent with observed data. The three theoretical pieces
  of predictions presented in the work, show that normal hierarchical
  model appears to be more favourable in nature than inverted hierarchical
  models.  
The perametrisation presented here is by no means unique but the
analysis presented here 
strengthens the foundation of $\mu$-$\tau$ symmetry in neutrino
sector,  based on realistic
GUT models.        

\section *{Acknowledgements}
NNS thanks the High Energy Physics Group, the Abdus Salam ICTP,
Trieste, Italy, for kind hospitality during the course of the work.

\end{document}